\documentclass[useAMS,usenatbib]{mn2e}
\usepackage{epsfig}
\usepackage{epstopdf}
\usepackage{graphicx}
\usepackage{amsmath}
\usepackage{amssymb}
\usepackage{natbib}
\usepackage[usenames]{color}
\newcommand{\kms}{\ensuremath{{\rm km\,s}^{-1}}}
\newcommand{\kpc}{\ensuremath{{\rm kpc}}}

\newcommand{\msun}{\ensuremath{{\rm M_{\odot}}}}
\newcommand{\lsun}{\ensuremath{{\rm L_{\odot}}}}
\newcommand{\hp}{\ensuremath{{h_{\rm p}}}}

\newcommand{\Sgas}{\Sigma_{\rm g}}
\newcommand{\dSstar}{\dot{\Sigma}_{\star}}

\newcommand{\sfrff}{{\rm SFR_{\rm ff}}}
\newcommand{\tff}{t_{\rm ff}}
\newcommand{\lbh}{L_{\rm BH}}
\newcommand{\lcii}{L_{\rm [CII]}}
\newcommand{\lco}{L_{\rm CO(1-0)}}
\newcommand{\lcoo}{L_{\rm CO(6-5)}} 
\newcommand{\Sdvco}{S\,{\Delta}v_{\rm CO(6-5)}}

\newcommand{\sfr}{{\rm SFR}}
\newcommand{\rvir}{R_{\rm vir}}
\newcommand{\rdisk}{R_{\rm d}}
\newcommand{\vcirc}{V_{\rm c}}

\newcommand{\fmol}{f_{\rm H_{2}}}
\newcommand{\fco}{f_{\rm CO}}

\newcommand{\mach}{\mathcal{M}}

\newcommand{\mhalo}{M_{\rm h}}
\newcommand{\mgas}{M_{\rm gas}}

\begin{document}

\title[Extreme Galaxies During Reionization]{Extreme Galaxies During Reionization: Testing ISM and Disk Models}

\author[Mu{\~n}oz and Furlanetto]
{
Joseph A.\ Mu{\~n}oz\thanks{E-mail: jamunoz@astro.ucla.edu} 
and 
Steven R. Furlanetto
\\Department of Physics and Astronomy, University of California Los Angeles; Los Angeles, CA 90095, USA
}

\maketitle

\begin{abstract}
We test the ability of equilibrium galactic disk and one-zone interstellar medium models to describe the physical and emission properties of quasar hosts, submillimeter galaxies, and Lyman-$\alpha$ emitters at $z\gtrsim6$.  The size, line widths, star formation rates, black hole accretion rates, gas masses and temperatures, and the relationships between these properties are all well-described by our model, and we provide approximate fitting formulae for comparison with future observations.  However, comparing our carbon line predictions to observations reveals differences between the ISM at low and high redshifts.  Our underestimate of the [CII] line emission indicates either higher star formation efficiencies in high-redshift molecular clouds or less depletion of metals into dust at fixed metallicity.  Further, our over-prediction of the CO(6--5)/CO(1--0) ratio suggests that molecular clouds in real high-redshift galaxies have a lower turbulent Mach number and more subthermal CO(6--5) emission than expected owing either to sizes smaller than the local Jeans mass or to a pressure support mechanism other than turbulence.
\end{abstract}


\section{Introduction}

Observations continue to probe the universe at ever higher redshifts with new discoveries of quasar hosts \citep[e.g.,][]{Fan00, Fan01, Fan03, Willott07, Willott09, Jiang09, Mortlock09, Willott10, Mortlock11}, sub-millimeter galaxies \citep[SMGs; e.g.,][]{Vieira13, Riechers13}, Lyman-break galaxies \citep[LBGs; e.g.,][]{Bouwens06, Bunker10, McLure10, Finkelstein10, Yan10, Bouwens11a, Bouwens11b, Oesch12, Coe13, Ellis13, Oesch13}, and Lyman-$\alpha$ emitters \citep[LAEs; e.g.,][]{Hu02a, Taniguchi05, Iye06, Ouchi09a, Ouchi09b, Ouchi10} at $z>6$.  These systems are different than those observed locally, with higher surface densities and smaller sizes besides being bathed in a significantly hotter Cosmic Microwave Background (CMB).  The more common LBGs and LAE, may also have much lower metallicities \citep[e.g.,][]{Finlator11, MF13a, Ouchi13}.  While observations of atomic and molecular lines have already begun to probe the state of the cool gas in these galaxies \citep[see][for a recent review]{CW13}, putting them into a complete theoretical framework for galaxy formation across cosmic time requires understanding the implications of these physical differences on the interstellar medium (ISM) at high-redshift \citep[e.g.,][]{Narayanan12, MF13a}.

Such a framework for galaxy formation includes the relationships among cosmic inflows, stellar and quasar outflows, gas transport through the galaxy, the buildup of the stellar population, and the growth of a central supermassive black hole in an active galactic nucleus (AGN).  We would also like to understand how the state of the ISM gas, the efficiency with which it forms stars, and the resulting molecular line emission relates to and can be a probe of the larger-scale physics.  In \citet{Munoz12}, we described the evolution of the galaxy luminosity function as a balance among star formation, star formation-driven outflows, and the gas accretion rate onto galaxies from the buildup of cosmological structure.  We then developed a model for galactic disks in \citet{MF12}---initially derived by \citet{Thompson05}---and embedded it into our cosmic setting to study the relationship between gas transport within galaxies and the growth of central black holes in the faintest high-redshift systems.  Finally, in \citet{MF13a}, we overlaid a photo-dissociative model for molecular clouds on sub-galactic scales to understand the chemical state of CO and make realistic predictions for ALMA and JVLA observations that self-consistently account for the physical conditions in the $z\sim6$ ISM.  Here we found that the CO signal from typical galaxies at these redshifts will be very difficult to observe if they are as metal- and dust-poor as expected, which both reduces the total amount of carbon and allows a higher fraction of the CO to be dissociated.

In the present work, we will continue to use this theoretical framework to describe galaxy formation during the reionization epoch.  While we previously focused on the CO signal from LBGs, we now turn our attention to CII---a product of CO dissociation---and its associated $158\,{\rm \mu m}$ line emission, which may be easier to detect than CO in low-metallicity environments.  We also apply our model to a broader range of $z=6$ galaxy types from LAEs---with relatively low star formation rates ($\lesssim 10\,{\rm \msun/yr}$)---to SMGs and quasar hosts---with star formation rates of tens to thousands of solar masses per year.  More data are available for these brighter systems \citep[e.g.,][]{Riechers09, Wang13, Willott13, Riechers13}, and the environments are undoubtedly extreme,making comparisons to our models more informative.  Our goals will be to ask whether our idealized framework for galaxy formation can describe SMGs and quasar hosts at all, to understand the physical sources of any disparity, and learn how to improve our treatment of sources currently undetected in molecular emission lines.

\subsection{Missing [CII] Emission}\label{sec:intro:cii}

One key aim of this work is to illuminate a puzzle in the relationship between molecular gas physics and the observed [CII] emission at high redshift.  Here we illustrate the difficulty with a simplified calculation (many of the parameter choices made here will be justified later on).  Let us assume that the host galaxies of the brightest $z=6$ quasars and SMGs live in $10^{13}\,\msun$ halos, that the cold gas in the galactic disk comprises 10\% of the halo's baryons,\footnote{If we assume that another 10\% of halo baryons are in stars, this results in a gas fraction---${\rm gas}/({\rm gas}+{\rm stars})$---of 50\%.} i.e., just over $10^{11}\,\msun$ worth, all of which has a temperature of $50\,{\rm K}$.  The carbon in some fraction, $\bar{f}_{\rm CO}$, of this gas is in the form of CO, while the rest is dissociated into CII.  The galactic disk, which extends out to 5\%/$\sqrt{2}$ of the halo virial radius, is about $3.5\,\kpc$ in this halo.  The average surface density of the disk is, thus, a few thousand solar masses per square parsec ($4200\,{\rm \msun/pc^2}$ for a cosmic baryon fraction of $f_{\rm b}=0.16$) and, given the redshift dependence of the halo virial radius, is expected to scale as $(1+z)^2$ for fixed mass.  This is well above the typical surface density of $85\,{\rm \msun/pc^2}$ observed in local clouds, so we assume that clouds in this galaxy have surface densities of order this higher value \citep{Krumholz09b}.  For a clumping factor of $c=5$, the chemical equilibrium calculation of \citet{Krumholz09b} predicts that such dense gas should be fully molecular unless the metallicity is below about $0.05\%$ of solar (below the threshold at which the equilibrium approximations in the model holds).  Moreover, at solar metallicity, the visual extinction through one of these clouds is about 200, and thus, approximately 99\% all of carbon gas is in the form of CO---sufficiently shielded against dissociation---according to the PDR model of \citet[][Eq. \ref{eq:fco}]{Wolfire10}.  50\% of the carbon is dissociated if the metallicity is approximately 1\% of solar.  

The maximum amount of [CII] emission is produced if the line is thermalized and optically thin.  In this optimistic case, the luminosity can be calculated simply by
\begin{equation}\label{eq:Lcii}
\lcii=\frac{\left(1-\bar{f}_{\rm CO}\right)\,\mgas\,{\rm C/H}}{m_{\rm p}}\,\frac{4\,{\rm e}^{-\hp\,\nu_{\rm [CII]}/k_{b}\,\bar{T}}}{2+4\,{\rm e}^{-\hp\,\nu_{\rm [CII]}/k_{b}\,\bar{T}}}\,A_{\rm [CII]}\,\hp\,\nu_{\rm [CII]},
\end{equation}
where $A_{\rm [CII]}=2.3\times10^{-6}\,{\rm s^{-1}}$, $\nu_{\rm [CII]}=1900.5\,{\rm GHz}$, $m_{\rm p}$ is the proton mass, and $\hp$ is the Planck constant.  We assume, moreover, that the carbon abundance is ${\rm C/H}=1.5\times10^{-4}$ at solar metallicity.  Then, a CO fraction of $\bar{f}_{\rm CO}=0.99$ gives a luminosity of about $0.5\times10^9\,\lsun$.  At high temperature, the ratio of the number of molecules in the excited state to the number in the ground state saturates at $2/3$ giving a luminosity of $1.4\times10^9\,\lsun$.  For comparison, the quasars J2310$+$1855 \citep{Wang13} and J1148$+$5251 \citep{Riechers09} and the submillimeter galaxy HFLS3 \citep{Riechers13} have [CII] luminosities of $8.8$, $26$, and $16\times10^{9}\,\lsun$, respectively.  These observations are an order-of-magnitude higher than the most optimistic cases in this  simple calculation.  Clearly, one of our seemingly reasonable assumptions does not extrapolate to these systems at $z\sim6$.  We plan to use our more sophisticated, physically motivated model of galaxy formation and line emission at high redshift to investigate this issue.

\subsection{This Paper}\label{sec:intro:outline}

We begin with a brief summary of the relevant details of the galaxy formation model developed in our previous work (\S\ref{sec:model}).  We then describe the observations of quasar hosts, SMGs, and LAEs that we take from the literature (\S\ref{sec:obs}) and interpret physical properties from these data that we will compare to our model.  We consider the suitability of our theoretical galactic disks for describing the observations (\S\ref{sec:comp}) with respect to their disk structure (\S\ref{sec:comp:struc}), quasar luminosities (\S\ref{sec:comp:lbh}), gas masses (\S\ref{sec:comp:mgas}), and gas temperatures (\S\ref{sec:comp:temp}).  Next, we turn our attention to the emission from carbon lines (\S\ref{sec:c}).  We compute the [CII] emission, presenting results using our fiducial model (\S\ref{sec:c:cii}) and considering additions (\S\ref{sec:c:alt}) that improve agreement with the [CII] data, before turning our attention to the line ratios of CO (\S\ref{sec:c:co}).  Finally, we conclude with a summary and discussion our results and their implications for high-redshift galaxy formation (\S\ref{sec:discussion}).  Throughout this work, we assume a flat, $\Lambda$CDM universe with $H_0=70\,{\rm km/s/Mpc}$, $\Omega_{\rm m}=0.28$, $\Omega_{\rm b}=0.046$, and $\sigma_8=0.82$.

\section{Galactic Disk Model}\label{sec:model}

We use the radiation pressure- and supernovae-supported disk models of \citet{Thompson05} and \citet{MF12} combined with the treatments of molecular clouds and carbon emission lines in \citet{MF13a}.  We refer the reader to these works for details of our model while only briefly summarizing our methods here.  We assume that gas accretes onto the outer edge of each disk at the cold flow rate \citep[e.g.,][]{McBride09} and, in our fiducial model, ignore suppression from AGN feedback in very large halo masses.  Star formation and winds deplete the gas as it moves toward the disk center.  The rate of star formation at each radius is set such that radiation pressure from starlight on dust and mechanical pressure from supernovae maintain marginal Toomre-instability (i.e., $Q=1$) and vertical hydrostatic equilibrium.\footnote{As in \citet{Thompson05}, we use opacities as functions of temperature and density derived from \citet{BL94} but have checked that the specific fits do not significantly affect our results.}  Meanwhile, we assume momentum driven super-winds eject gas at a rate $(400\,{\rm km/s})/\sigma$ times higher than the star formation rate, where $\sigma$ is the halo velocity dispersion, matching observations of the UV luminosity function at $z=6$--$8$ \citep{ML11, Munoz12} and consistent with the results from numerical simulations \citep{OD08}.\footnote{While \citet{Munoz12} determined this mass-loading factor for LBGs at $z=6$--$8$, we will also apply it here to quasar hosts whose metallicities, stellar populations, and environments may be somewhat different.  However, given the large host halo masses of these systems and their correspondingly high velocity dispersions, these stellar winds contribute only minimally to gas depletion so that our assumption does not substantially affect the calculation.}  We consider two phenomenological models for the transport rate of gas through the disk: a linear spiral wave (LSW) model in which the inflow velocity is $v_{\rm in}=m\,c_{\rm s}$ and a shocked, nonlinear inflow model where $v_{\rm in}=\sqrt{2}\,\beta\,\sigma$.  Here, $c_{\rm s}$ is the local sound speed and the two free-parameters defining these prescriptions are the Mach number, $m$, and the constant $\beta$.

We further assume that the gas remaining after depletion by star formation and winds transitions smoothly into the accretion disk of a central black hole and powers an AGN.  In LBGs, the black hole growth rate is negligible compared to the total star formation rate, and while the resulting X-rays dominate the contribution from high-mass X-ray binaries in the stellar disk, they are currently undetectable in stacked samples of {\it{Chandra}} data \citep{MF12}.  However, with high halo masses and for very rapid gas inflow, e.g., $\beta$-models with $\beta \gtrsim 0.01$, the black hole growth rate can be significant enough to power an observable quasar.  

At each radius in the disk, we then assume that the gas fragments into clouds on the scale of the local Jeans mass.  Calculating the fraction of molecular gas turned into stars, $\sfrff$, per free-fall time, $\tff$, as prescribed by \citet{KM05}, we set the fraction of cloud gas in molecular form, $\fmol$, to be such that the required star formation rate is produced.  That is,
\begin{equation}\label{eq:fmol}
\fmol=\frac{\dSstar}{\Sgas}\,
\left( \frac{\sfrff}{\tff}\right)^{-1},
\end{equation}
where $\dSstar$ and $\Sgas$ are the surface star formation rate and gas densities as a function of galactocentric radius.  Additionally, where $\fmol$ would be greater than unity, we simply set $\fmol=1$.\footnote{Imposing the threshold of $\fmol=1$ effectively increases the star formation efficiency of the gas beyond that prescribed by \citet{KM05}.  Physically, this may result from gas out of chemical equilibrium in which star formation is produced in atomic gas \citep[e.g.,][]{Krumholz12}.}  Thus, in general, more efficient star formation results in lower molecular fractions since the dynamically-balanced star formation rate is unchanged.  We then use the \citet{Wolfire10} model of cloud photo-dissociated regions (PDRs) to calculate the fraction, $\fco$, of cloud mass in which carbon is in the form of CO rather than dissociated CII, where $\fco \leq \fmol$.  CO in the clouds are shielded from the external dissociating radiation field, $G_0$, by turbulently-generated inhomogeneities that follow a log-normal distribution where the ratio of the median density to the average is $\sqrt{1+3\,\mach^2/4}$, with $\mach$ is the thermal Mach number of the turbulent gas.  \citet{Wolfire10} models these inhomogeneities as uniform clumps of density $n_{\rm c}$ embedded within smooth, more diffuse gas.  The ratio $\fco/\fmol$ is then given by
\begin{equation}\label{eq:fco}
{\ln}\left(\frac{\fco}{\fmol}\right)=\frac{-4.0}{A_{V}}\,\left[0.53-0.045\,{\ln}\left(\frac{G'_0}{n_{\rm c}\,{\rm cm^3}}\right)-0.097\,{\ln}(Z')\right],
\end{equation}
where $G'_0=G_0/2.7\times10^{-3}\,{\rm erg\,s^{-1}\,cm^{-2}}$ is the external far-UV radiation field bathing the cloud in units of the local Galactic interstellar field found by \citet{Draine78}, $Z'$ is the metallicity in solar units, and we set $n_{\rm c}$ to be the median gas density in the cloud.  The far UV dissociating radiation field is determined produced primarily by starlight, and we have checked that emission from the AGN in our model galaxies contributes negligibly.

\begin{table*}
\begin{center}
\caption{ Derived Physical Properties }
\begin{tabular}{lccccccc}
\hline
Name & $z$ & $R_{\rm disk} $ & $v_{\rm c}$ & sin $i$ & SFR & Type & Refer. \\ 
& & kpc & km/s & & ${\rm \msun/yr}$ & & \\ 
(1) & (2) & (3) & (4) & (5) & (6) & (7) & (8) \\ 
\hline
\hline
SDSS J231038.88$+$185519.7 & 6.00 & 3.18$\pm$0.17 & 549$\pm$46 & 0.71 & 2100* & QSO & A \\
ULAS J131911.29$+$095051.4 & 6.13 & 3.26$\pm$0.39 & 620$\pm$143 & 0.84 & 1400* & QSO & A \\
SDSS J205406.49$-$000514.8 & 6.04 & 1.98$\pm$0.28 & 589$\pm$55 & 0.41 & 380* & QSO & A \\
SDSS J012958.51$-$003539.7 & 5.78 & 2.41$\pm$0.35 & 242$\pm$31 & 0.80 & 40* & QSO & A \\
SDSS J104433.04$-$012502.2 & 5.78 & 3.54$\pm$0.64 & 160$\pm$60 & --\,-- & 2800* & QSO & A \\
SDSS J114816.64$+$525150.3 & 6.42 & 4.93$\pm$0.88 & 287$\pm$28 & --\,-- & 3300 & QSO & B \\
CFHQS J021013$-$045620 & 6.43 & 2.85$\pm$1.36 & 189$\pm$18 & --\,-- & 48$\pm$14 & QSO & C \\
CFHQS J232908$-$030158 & 6.42 & --\,-- & --\,-- & --\,-- & $< 40$ & QSO & C \\
1HERMES S350 J170647.8$+$584623 & 6.34 & 3.4 & 470$\pm$135 & --\,-- & 2900 & SMG & D \\
{\it Himiko} & 6.60 & $\sim 16$ & --\,-- & --\,-- & $\sim 100$ & LAE & E, F \\
{\it IOK$-$1} & 6.96 & $< 5.35$ & --\,-- & --\,-- & $\sim 16$ & LAE & F \\
HCM 6A & 6.56 & --\,-- & --\,-- & --\,-- & $9$ & LAE & G \\
\hline
\end{tabular}
\begin{list}{}{}
\item{Col. (1): Source name.  Col. (2): Redshift.  Col. (3): Disk radius.  Col. (4): Disk circular velocity.  Col. (5): Disk inclination.  Col. (6): Star formation rate.  Col. (7): Reference: (A) \citet{Wang13}; (B) \citet{Riechers09}; (C) \citet{Willott13}; (D) \citet{Riechers13}; (E) \citet{Ouchi13}; (F) \citet{Walter12}; (G) \citet{Hu02b}.  *Predicted values based on Eq. \ref{eq:fits} with $\beta=0.1$.}
\end{list}
\label{tab:disk}
\end{center}
\end{table*}

The temperature of the gas in the absence of the CMB is given by the disk model.  We combine this value with the CMB temperature at high redshift as in \citet{MF13a} assuming that all of the CMB radiation is ultimately transferred to the gas.  The temperature and density of each cloud is then input into a version of the escape-probability formalism code by \citet{KT07}, where the modifications, described in \citet{MF13a}, are consistent with those implemented in the new DESPOTIC code \citep{Krumholz13}.  Given the lack of observational constraints at high redshift, this procedure requires fewer arbitrary parameter choices than does the detailed heating and cooling balance performed by DESPOTIC and assumes a regime in which the dust and gas are tightly coupled.  However, since our temperatures are consistent with those determined from observations of emission lines in high-z quasar hosts (see \S\ref{sec:comp:temp}), we conclude that our method is sufficient for our purposes.  The escape-probability code includes the full log-normal density distribution to determine the CO and CII level populations and calculates the resulting line emission from each cloud.  Throughout the rest of this Paper, we assume a carbon abundance of ${\rm C/H}=1.5\times10^{-4}\,Z'$.

In considering only the emission from the PDRs of cold molecular clouds, we have ignored any contribution from HII regions, which are not well-described by our model.  These ionized regions would both lower the molecular fraction of the galactic gas and increase its temperature.  However, observational and theoretical evidence suggest that this is likely a small effect.  As we will see in \S\ref{sec:comp:temp}, the cold gas temperatures in our models without HII regions are comparable to those inferred from both dust and molecular line observations of quasar hosts suggesting that the hotter, ionized gas in HII regions does not contribute significantly.  Moreover, in a sample of 60 normal, star-forming galaxies at relatively low redshift, \citet{Malhotra01} attributed only about 50\% of the total [CII] flux to ionized gas as probed by [NII], while \citet{Vasta10} found somewhat less emission in more recent work.  We would expect HII regions in molecular clouds to be even less important in starbursts or AGN or at much higher redshifts, at least {\it{dynamically}}, as external pressure from the ISM becomes significant \citep{Krumholz06}.  Additionally, the assumption of no contribution from HII regions is consistent with the lack of a [NII] detection in J1148$+$5251 \citep{Walter09a}.  Despite this evidence, in \S\ref{sec:c:alt}, we will consider the effect of an alternative addition to our model in which the molecular fraction is lower than computed by equation \ref{eq:fmol}, which is similar to the effect of including a contribution from HII regions, albeit at lower temperatures.

Finally, we arrive at the observed signal by summing the emission from all clouds in our model galaxies and subtracting the CMB as an observational background.  We note that, while our model accurately determines the intrinsic cloud emission as well as the effect of the CMB background, in some cases, the two are of similar magnitude and the subtracted result is not well-determined \citep{MF13a}.  This is particularly true for the CO(1--0) line of galaxies in low-mass halos but is not an issue for the [CII] line in the present work.

\section{Observed Systems}\label{sec:obs}

\begin{table*}
\begin{center}
\caption{ Summary of Observed Emission }
\begin{tabular}{lcccccc}
\hline
Shortened Name & $L_{\rm BH}$ & $\lcii$ & $\lco$ & $\Sdvco$ & Type & Refer. \\ 
& $10^{13}\,{\rm \lsun}$ & $10^{9}\,{\rm Jy\,km/s\,Mpc^2}$ & $10^{5}\,{\rm \lsun}$ & Jy km/s & & \\ 
(1) & (2) & (3) & (4) & (5) & (6) & (7) \\ 
\hline
\hline
J2310$+$1855 & 9.3 & 64$\pm$9.0 & 32.1\dag & 1.52$\pm$0.13 & QSO & A \\
J1319$+$0950 & 7.0 & 28$\pm$6 & 9.4\dag & 0.43$\pm$0.09 & QSO & A \\
J2054$-$0005 & 2.8 & 22$\pm$3 & 7.3\dag & 0.34$\pm$0.07 & QSO & A \\
J0129$-$0035 & 0.57 & 12$\pm$2 & 7.4\dag & 0.37$\pm$0.07 & QSO & A \\
J1044$-$0125 & 11.6 & 10$\pm$3 & 4.2\dag & 0.21$\pm$0.04 & QSO & A \\
J1148$+$5251 & 10 & 165$\pm$13 & $< 7100$ & 0.67$\pm$0.08 & QSO & B \\
J0210$-$0456 & 0.53 & 1.94$\pm$0.265 & --\,-- & --\,-- & QSO & C \\
J2329$-$0301 & 1 & $< 0.71$ & --\,-- & --\,-- & QSO & C \\
HFLS3 & --\,-- & 100.$\pm$21 & 4800$\pm$1500 & 2.74$\pm$0.68& SMG & D \\
{\it Himiko} & --\,-- & $< 0.33$ & --\,-- & --\,-- & LAE & E \\
{\it IOK$-$1} & --\,-- & $< 2.86$ & --\,-- & --\,-- & LAE & F \\
{\it HCM 6A} & --\,-- & $< 0.41$ & --\,-- & --\,-- & LAE & G \\
\hline
\end{tabular}
\begin{list}{}{}
\item{Col. (1): Shortened source name.  Col. (2): Bolometric black hole luminosity.  Col. (3): [CII] line luminosity.  Col. (4): CO(1--0) line luminosity.  Col. (5): CO(6--5) velocity-integrated line flux.  Col. (6): Source type.  Col. (7): Reference: (A) \citet{Wang13}; (B) \citet{Willott03} and \citet{Riechers09}; (C) \citet{Willott13}; (D) \citet{Riechers13}; (E) \citet{Ouchi13}; (F) \citet{Walter12}; (G) \citet{Kanekar13}.  \dag Calculated by \citet{Wang13} from CO(6--5) observations assuming an excitation ladder similar to that of J1148$+$5251 \citep{Riechers13}.}
\end{list}
\label{tab:lum}
\end{center}
\end{table*}

In this section, we discuss the observed systems to which we will compare our model.  These sources fall into three categories: (1) quasar hosts (QSOs), (2) sub-mm galaxies (SMGs), and (3) Lyman-$\alpha$ emitters (LAEs).  Since the redshift of the system is an {\it a priori} requirement for observing the [CII] line flux (at least given current instruments), Lyman-break galaxies with no discernible line emission are not as good candidates for this type of followup at high-redshift.  We plan to extend our analysis to these objects using the aggregate, unresolved intensity on the sky in forthcoming work.

Table \ref{tab:disk} shows the sources we have compiled from the literature and lists, where available, their derived physical properties: size, circular velocity, inclination, and star formation rate, along with a designation of the type of system.  We have interpreted each object as if it were a galactic disk.  We determined the disk radius for the quasar hosts and SMGs by taking the semi-major axis of the [CII] emission on the sky (where detected) except in the case of J1148$+$5251, whose more extended CO emission clearly traces galactic disk gas.  Sizes for LAEs were derived from the extent of the Lyman-$\alpha$ emission, but we do not interpret these as disk radii since the Lyman-$\alpha$ results from scattering of material outside of the galactic disk.  We computed the  angular diameter distance based on the specific redshift of each object, where an angular size of 0.1" corresponds to about $5.7\,{\kpc}$ at $z=6$.

We relate the circular velocity, $\vcirc$, to the [CII] FWHM line width, ${\Delta}v_{\rm [CII]}$, by assuming $\vcirc={\Delta}v_{\rm [CII]}$ \citep{Nagamine06b}.  We use this relation in the absence of other inclination information and set $\vcirc={\Delta}v_{\rm [CII]}/{\rm sin}\,i$, otherwise.  Quoted errors in $\vcirc$ combine contributions from both ${\Delta}v_{\rm [CII]}$ and ${\rm sin}\,i$ where available.

The star formation rates from our sample were typically derived by assuming a ratio with observed FIR luminosity.  However, no such measurements or derived star formation rates are yet available for the \citet{Wang13} systems.  Of particular difficulty is the star formation rate of HCM 6A \citep{Hu02a}, which varies wildly when inferred from different observational indicators \citep[see][for a summary]{Kanekar13}.  For this object, we assume the star formation rate inferred from UV continuum measurements of roughly $9\,{\rm \msun/yr}$ \citep{Hu02b}, which ignores the effect of dust obscuration.

Table \ref{tab:lum} lists, for these same systems, their luminous properties: bolometric black hole luminosity, [CII] luminosity, CO(1--0) luminosity, and CO(6--5) flux.  Typically, the bolometric black hole luminosities were derived from measurements in the rest-frame UV assuming a bolometric correction.\footnote{\citet{Wang13}, for example, uses $\lbh=4.2\,\nu\,L_{\rm 1450}$, the simplest fit for the isotropic correction from \citet{Runnoe12}.}  Direct information about the CO(1--0) line is not actually available for the sample of quasars from \citet{Wang13}, but these authors have assigned a value to each object based on the CO(6--5) data assuming a model excitation ladder similar to that of J1148$+$5251 \citep{Riechers09}.  However, the extrapolation of high-order CO lines to lower orders is quite uncertain \citep[e.g.][]{CW13}.

Particularly pertinent to comparison with our models is the presence (or lack thereof) of an AGN in HFLS3.  The absence of AGN activity was determined from the galaxy's lack of excess radio emission, hot dust component, rest-frame optical continuum emission up to $2.4\,{\rm \mu m}$ beyond that associated with the starburst, and broad rest-frame UV lines \citep[][and private communication with the authors]{Riechers13}.  However, these criteria may not be sufficient to completely rule out the presence of a weak AGN \citep{Rush96, Spinoglio02, EM12}.  If HFLS3 does indeed have such a faint nuclear component, we can ask at what level such a component would remain undetected.  Combining the measured 2.2$\,\mu$m and 3.6$\,\mu$m emission with the recommended bolometric correction fits of \citet{Runnoe12}, we conservatively estimate the bolometric black hole luminosity to be less than about $3\times10^{12}\,\lsun$.  For comparison, J0129$-$0035 was detected in an extremely deep stripe of SDSS data with a rest-frame UV magnitude of 22.28 \citep{Jiang09}---well-below the typical SDSS-DR9 $z$-band detection limit of 20.5---corresponding to an isotropic bolometric black hole luminosity of $5.7\times10^{12}\,\lsun$.  We nominally set a maximum luminosity of $3\times10^{12}\,\lsun$ on an AGN component of HFLS3.

Where necessary, we have assumed a maximum black hole luminosity for the LAEs in our sample of $10^{12}\,\lsun$, small enough that the rest-frame UV emission in these systems is still dominated by starlight.

\section{Comparison of Model Galactic Disks to Observations}\label{sec:comp}

In this section, we compare our model galactic disks to observed systems with respect to their disk structure and dynamics (\S\ref{sec:comp:struc}), quasar luminosities (\S\ref{sec:comp:lbh}), gas masses (\S\ref{sec:comp:mgas}), and gas temperatures (\S\ref{sec:comp:temp}).

\subsection{Structure and Dynamics}\label{sec:comp:struc}

\begin{figure}
\begin{center}
\includegraphics[width=\columnwidth]{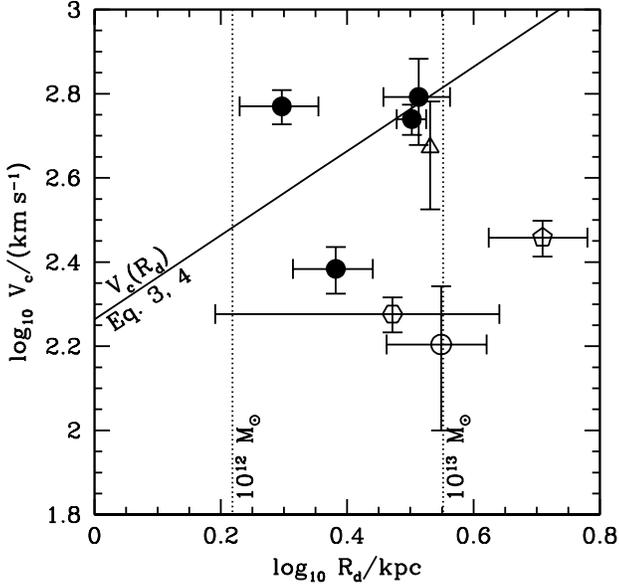}
\caption{\label{fig:disk}
Disk circular velocity, $\vcirc$, as a function of disk radius, $\rdisk$, for quasar hosts and SMGs at $z\sim6$.  The solid line indicates our model derived from Eq. \ref{eq:rdisk} and \ref{eq:vcirc} at $z=6$, while vertical, dotted lines denote disk radii corresponding to $10^{12}$ and $10^{13}\,\msun$.  The points show observations from Table \ref{tab:disk} where available.  Circles correspond to the quasar sample of \citet{Wang13}, the pentagon is J1148$+$5251, the hexagon is J0210$-$0456, and the triangle is the SMG HFLS3.  Filled points indicate observations corrected for inclination.
}
\end{center}
\end{figure}

Here, we compare the structure and dynamics of our model disks to those observed and show that our simple framework describes the data reasonably well.  We first consider disk radii by assuming a constant spin parameter of $\lambda=0.05$.  The outer radius of the disk is then given by
\begin{equation}\label{eq:rdisk}
\rdisk=\frac{\lambda\,\rvir}{\sqrt{2}}=1.65\,\kpc\,\left(\frac{\mhalo}{10^{12}\,\msun}\right)^{1/3}\,\left(\frac{1+z}{7}\right)^{-1},
\end{equation}
where $\rvir$ is the virial radius of a dark matter halo with mass $\mhalo$ at redshift $z$.  The circular velocity of a model disk is taken to be 
\begin{equation}\label{eq:vcirc}
\vcirc=\sqrt{2}\,\sigma=303\,\kms\,\left(\frac{\mhalo}{10^{12}\,\msun}\right)^{1/3}\,\left(\frac{1+z}{7}\right)^{1/2},
\end{equation}
where $\sigma$ is the halo velocity dispersion \citep{BL01}.  Inspection of equations \ref{eq:rdisk} and \ref{eq:vcirc} indicates that the ratio of $\vcirc$ to $\rdisk$ will be independent of mass and fixed at a given redshift.  Figure \ref{fig:disk} shows this relationship at $z=6$.  

\begin{figure}
\begin{center}
\includegraphics[width=\columnwidth,trim=40 30 220 20,clip]{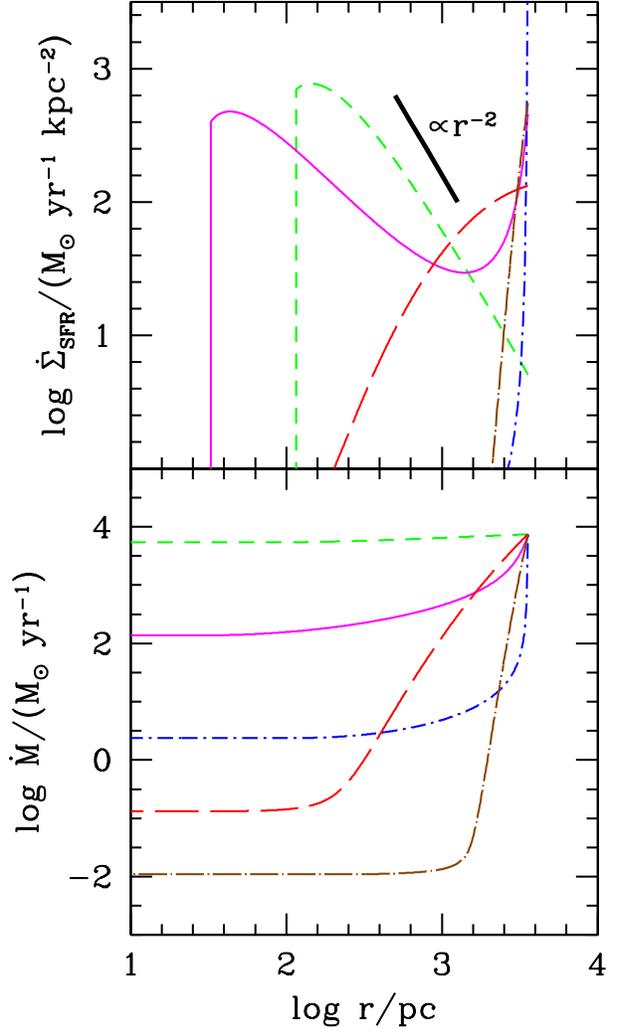}
\caption{\label{fig:rdist}
The star formation rate surface density (top panel) and mass inflow rate (bottom panel) as a function of galactocentric radius, $r$, in our disk model for a host halo mass of $10^{13}\,\msun$ at $z=6$.  Results are plotted out to the disk radius of approximately $3.55\,\kpc$.  We show the behavior of a range of angular momentum transport models: short-dashed (green), solid (magenta), and dot-short-dashed (blue) lines denote $\beta$-models with $\beta=1$, 0.1, and 0.01, while long-dashed (red) and dot-long-dashed (brown) curves indicate LSW models with $m=1$ and 0.02, respectively.  The thick, solid line in the upper panel references a proportional relationship with $r^{-2}$.
}
\end{center}
\end{figure}

Our model disk sizes and circular velocities appear to describe some of the objects in our sample very well.  Systems for which inclinations are available (filled symbols) are particularly well-represented.  On the other hand, objects for which we do not have inclinations (open symbols) are also those most off-set from our model relation.  Moreover, these objects are all below the relation so that incorporating an inclination would likely improve agreement in each case.  Additionally, we note that the errors on the size of J0210$-$0456 (hexagon) are particularly large making a comparison difficult.  The vertical, dotted lines in the figure indicate disk radii for halo masses of $10^{12}$ and $10^{13}\,\msun$ at $z=6$ according to equation \ref{eq:rdisk}.  Judged purely  on inferred disk radius (to avoid the additional complication of missing inclinations), the objects in our sample all have halo masses roughly in this range.

The nature of our models allows us to consider the disk structure in even finer detail.  Figure \ref{fig:rdist} plots the radial distribution of star formation and gas inflow in our model for a $10^{13}\,\msun$ halo at $z=6$ and a range of angular momentum transport models.  We specifically consider the range of transport models summarized in Table \ref{tab:amt}.  Models with faster inflow rates of gas maintain higher star formation rates into the inner regions of the galactic disk \citep[see also][]{Thompson05, MF12, MF13a}.  Additionally, for $\beta \gtrsim 0.1$, $\dSstar$ peaks at a radius of $\sim100\,{\rm pc}$ reaching nearly $10^{3}\,{\rm \msun/yr/kpc^2}$.  This peak is qualitatively similar to the extreme nuclear star formation rate densities noted in HFLS3---$\dSstar\sim600\,{\rm \msun/yr/kpc^2}$ distributed over a $1.3\,\kpc$ radius region \citep{Riechers13}---and J1148$+$5251---$\sim10^3\,{\rm \msun/yr/kpc^{2}}$ over $\sim750\,{\rm pc}$ \citep{Walter09b}---two of the most massive systems in our sample as determined by galactic radius.  This behavior in our models is not a consequence of the opacity gap discussed in \citet{Thompson05} and appears even when the opacity is set to be constant.  Rather, it results simply from a mass inflow rate and inflow velocity that are roughly constant with radius.  As a result, the disk scale height $h \propto r$ and the star formation surface density required to maintain marginal Toomre-instability and vertical hydrostatic equilibrium is $\dSstar \propto r^{-2}$.  The star formation rate turns over at very small radii as thermal pressure from the gas becomes more important than stellar feedback \citep{Thompson05, MF12}.

As radius decreases toward the center of the disk, once star formation is no longer necessary for the stability of the disk, its value drops, no further gas is depleted by star formation or winds, and the mass inflow rate becomes constant.  The behavior of constant inflow rate at small radii is shown in the lower panel of Figure \ref{fig:rdist}.  Our model assumes that all remaining inflowing gas is accreted onto the central black hole and powers an AGN.  

\begin{figure*}
\begin{center}
\includegraphics[width=\textwidth,trim=0 350 0 0,clip]{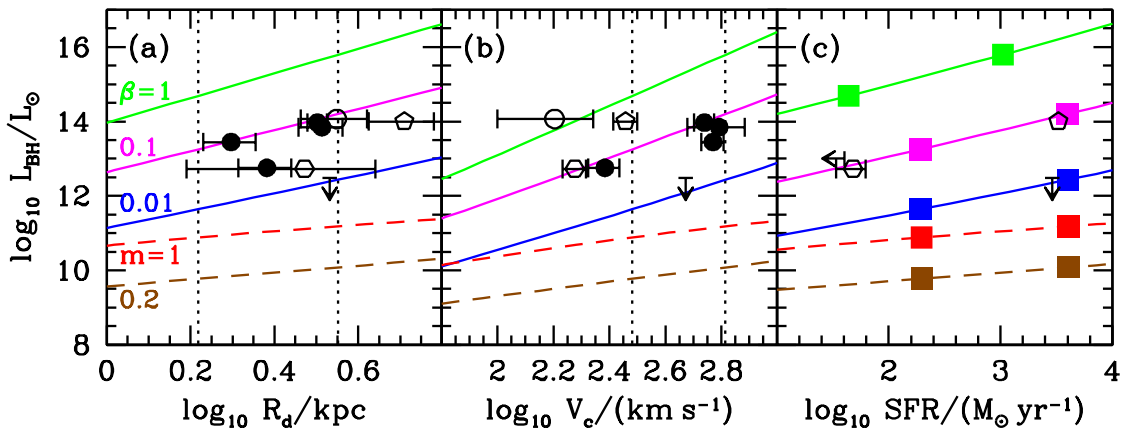}
\caption{\label{fig:lbh}
Bolometric black hole luminosity at $z\sim6$ as a function of $\rdisk$ (panel a), $\vcirc$ (panel b), and galactic star formation rate (panel c).  Solid lines denote results for $\beta$-models with $\beta=1$ (green), 0.1 (magenta), and 0.01 (blue), from top to bottom, while dashed curves indicate results for LSW models with $m=1$ (upper, red) and 0.02 (lower, brown).  Each point along our model lines represents a different host halo mass with $10^{12}$ and $10^{13}\,\msun$ denoted by by vertical dotted lines in panels (a) and (b) and filled squares in panel (c).  Note that models with less black hole accretion than that for $\beta=0.1$ all have the same relationship between star formation rate and halo mass.
}
\end{center}
\end{figure*}

\begin{table}
\begin{center}
\begin{tabular}{lcc}
\hline 
\multicolumn{3}{c}{Angular Momentum Transport Models} \\ 
\hline
model name & parameter & possible galaxy types \\
\hline
$\beta$-model & $\beta=0.01$ & QSOs, SMGs \\
$\beta$-model & $\beta=0.1$ & QSOs, SMGs \\
$\beta$-model & $\beta=1$ & QSOs, SMGs \\
LSW & $m=0.2$ & LBGs, LAEs \\
LSW & $m=1$ & LBGs, LAEs \\
\hline
\end{tabular}
\caption{The range of angular momentum transport model that we consider in this work.  See \S\ref{sec:model} for a description of the implementation of each model.
}
\label{tab:amt}
\end{center}
\end{table}

The consistency between our simple disk model and the structural features and relationships of high-redshift quasars is somewhat surprising given the picture of these systems as the product of rapid mergers and disruptions \citep[e.g.,][]{Li07}.  However, this agreement supports our continued use of the model to describe quasar luminosities, star formation rates, and molecular line luminosities in the subsequent sections.

\subsection{Quasar Luminosities}\label{sec:comp:lbh}

Next, we consider the luminosity of central black holes in our model galaxies and the ability of this framework to describe the relationships between observed $z\sim6$ quasars and their host properties.  Figure \ref{fig:lbh} plots the bolometric luminosities of our central black holes, $\lbh$, versus $\rdisk$, $\vcirc$, and the total star formation rate, respectively.  Solid curves assume that the disk gas inflow is non-linear (i.e. for a $\beta$-model), while dashed curves show results for an LSW disk.  Table \ref{tab:amt} summarizes the range of angular momentum transport models that we consider in this work.  We find that using the optically thick dust model of \citet{Thompson05} produces negligible differences from the dust-free model of \citet{MF12} for rapid angular momentum transport, but the two begin to deviate for models where the inflow rate is slower with the dusty model gives somewhat more black hole growth than in the dust-free case if one assumes an LSW disk.  The figure shows that, as expected, the faster transport of angular momentum in the $\beta$-model produced significantly more accretion onto the central black hole than does that for the LSW models for the parameter ranges we considered.  Here, we present fits to our model results to aid in future predictions.\footnote{We define ``fits" loosely here and without any statistical rigor.  Our goal is merely to describe our results for interpolation purposes rather than to provide physical insight.}  The fits are described by the form
\begin{equation}\label{eq:fits}
y/y_0=\zeta^{a}\,\left(\frac{x}{x_0}\right)^{b+c\,{\rm log}_{10}\,\zeta},
\end{equation}
with $y/y_0$, $x/x_0$, $a$, $b$, and $c$ given in Table \ref{tab:fits}.  For nonlinear $\beta$-models, $\zeta=\beta$, while for LSW models, $\zeta=m$.

\begin{table*}
\begin{center}
\begin{tabular}{lccccc}
\hline
Inflow model & $y/y_0$ & $x/x_0$ & $a$ & $b$ & $c$ \\ 
\hline
$\beta$-model & $\lbh/10^{14}\,\lsun$ & $\rdisk/\kpc$ & 1.5 & 3.3 & 0.4 \\
$\beta$-model & $\lbh/1.3\times10^{13}\,\lsun$ & $\vcirc/100\,{\rm km\,s^{-1}}$ & 1.5 & 3.3 & 0.4 \\
$\beta$-model & $\lbh/2.4\times10^{13}\,\lsun$ & ${\rm SFR}/{\rm \msun\,yr^{-1}}$ & 1.68 & 0.8 & 0.086 \\
LSW & $\lbh/4.5\times10^{10}\,\lsun$ & $\rdisk/\kpc$ & 1.56 & 0.9 & 0.0 \\
LSW & $\lbh/2.3\times10^{10}\,\lsun$ & $\vcirc/100\,{\rm km\,s^{-1}}$ & 1.56 & 0.98 & 0.0 \\
LSW & $\lbh/2.0\times10^{10}\,\lsun$ & ${\rm SFR}/{\rm \msun\,yr^{-1}}$ & 1.57 & 0.18 & 0.0 \\
\hline
\end{tabular}
\caption{Approximate fits to our model results based on the fitting function in Eq. \ref{eq:fits}, where $\zeta=\beta$ in non-linear inflow models and $\zeta=m$ in LSW models.}
\label{tab:fits}
\end{center}
\end{table*}

For a $\beta$-model, we find that $\lbh$ is nearly proportional to halo mass.  Moreover, for $\beta>0.1$, the inflow rate becomes high enough that the amount of gas bypassing star formation to fuel the central black hole becomes significant, and the star formation rate loses its proportionality with halo mass.  This effect is shown in panel (c) of Figure \ref{fig:lbh}, where the squares indicating $10^{12}$ and $10^{13}\,\msun$ halos are at lower star formation rates for $\beta=1$ than are those for $\beta=0.1$.  While the lower inflow velocities associated with LSW models result in dependences on $\rdisk$, $\vcirc$, and star formation rate that are not quite as well-fit by power-laws as those for a $\beta$-model, we nevertheless attempt to describe the behavior as such for interpolation purposes.  However, we caution against using these fits to extrapolate beyond the range plotted in Figure \ref{fig:lbh}.  We find that $\lbh$ is a much shallower function of halo mass ($\propto \mhalo^{\sim 1/3}$) than for $\beta$-models, which implies that the BH fueling is not supply-limited.  Note that we obtain better descriptions of the results for each independent variable by fitting each directly from the plots in Figure \ref{fig:lbh} than by assuming any knowledge of the true relationships between the variables (for example, by using equations \ref{eq:rdisk} and \ref{eq:vcirc}) owing to the different parameter ranges over which we fit the results.

As can be seen in Figure \ref{fig:lbh}, the relationships between $L_{\rm BH}$ and the disk parameters of $\rdisk$, $\vcirc$, and star formation rate for observed quasar hosts are well-matched by a $\beta$-model disk with $\beta$ slightly less than 0.1 and halo masses in the range of $10^{12}$--$10^{13}\,\msun$.  These sources produce black hole luminosities well in excess of that expected in any of the LSW models, which appear to be ruled out as descriptions for quasar hosts.  This is in contrast to more typical galaxies, such as LBGs and LAEs, which can only be represented by $\beta$-model systems if a significant amount of the resulting black hole emission is obscured \citep{MF12}.  Given the agreement between our model and the data, we use equation \ref{eq:fits} and the fit parameters in Table \ref{tab:fits} with $\beta=0.1$ to predict star formation rates for the \citet{Wang13} sample of quasars based on their bolometric black hole luminosities and list these predictions in Table \ref{tab:disk} for comparison with upcoming data.

While Figure \ref{fig:disk} showed several disagreements between our sample and our model $\vcirc$-$\rdisk$ relation, particularly for J1044$-$0125 and J1148$+$5251, we note that the relationship between $\lbh$ and $\rdisk$ for these systems is generally consistent with that of the other quasars in Panel (a) of Figure \ref{fig:lbh}.  Consequently, we expect that the discrepancies in Figure \ref{fig:disk}---and consequently in panel (b) of Figure \ref{fig:lbh}---indeed owe simply to an under-estimated $\vcirc$ resulting from a inclined disks.

Interestingly, both CFHQS quasars that we consider---J0210$-$0456 (hexagon) and J2329$-$0301 (upper limit on star formation rate in panel c)---are consistent with our model if $\beta=0.1$, though the radius of J0210$-$0456 may be somewhat overestimated---not completely unexpected given the large errors on the observed size.  While \citet{Willott13} suggested that J2329$-$0301, in particular, may be observed in a rare state of quasar-suppressed star formation, we obtain agreement with our model with no such feedback mechanism.  J2329$-$0301 appears to have a smaller star formation rate simply because it is hosted in a smaller system 

Given its apparent lack of quasar activity, HFLS3 (upper limit on $\lbh$) is inconsistent with a $\beta$-model with $\beta>0.01$.  However, assuming the data allow an undetected bolometric luminosity of $3\times10^{12}\,\lsun$, we cannot rule out a $\beta$-model description of HFLS3 with $\beta=0.01$.

Of course, a variety of uncertainties in both our models and in the observations may cause shifts in the plotted quantities.  For example, our model assumes that all gas not turned into stars or ejected by stellar radiation-pressure driven winds accretes onto the central black hole and that all of the resulting radiation escapes.  Thus, our model black hole luminosities may be over-estimates (1) if angular momentum is not shed quite so quickly in the central portions of the disk, resulting in the buildup of a bulge \citep[e.g.][]{Dekel09b} rather than direct accretion onto the black hole, (2) if a wind from the central AGN---inferred in some observations by a broad component of [CII] \citep{Maiolino12, Valiante12} and not included in our model---blows out gas from the center that would otherwise be accreted, (3) if AGN feedback results in a reduced gas accretion rate onto the galactic disk \citep[e.g.,][]{Voort11}, or (4) if radiation trapping \citep[e.g.,][]{WL11b} or obscuration prevents some of the emission from the accreted gas from escaping.  Including a significant contribution from any of these effects would result in lower values of $\lbh$ in our model and would require a larger value of $\beta$ to reproduce the observations.  At the same time, the bolometric corrections used by \citet{Wang13} are subject to their own uncertainties and extrapolations, and the total luminosity is over-estimated if the emission is not isotropic \citep{Runnoe12}.  In this latter case, the observations could be better described by our models with a lower value of $\beta$.  Additionally, the relationship between FIR luminosity and star formation rate is only empirically calibrated at low redshift and may not hold at $z\sim6$.  Ultimately, however, if any of the above effects are significant, they would all have to conspire fairly precisely to produce the agreement we find between most of the quasars in our sample and a disk model with fixed $\beta$.

\subsection{Gas Masses}\label{sec:comp:mgas}

\begin{figure}
\begin{center}
\includegraphics[width=\columnwidth]{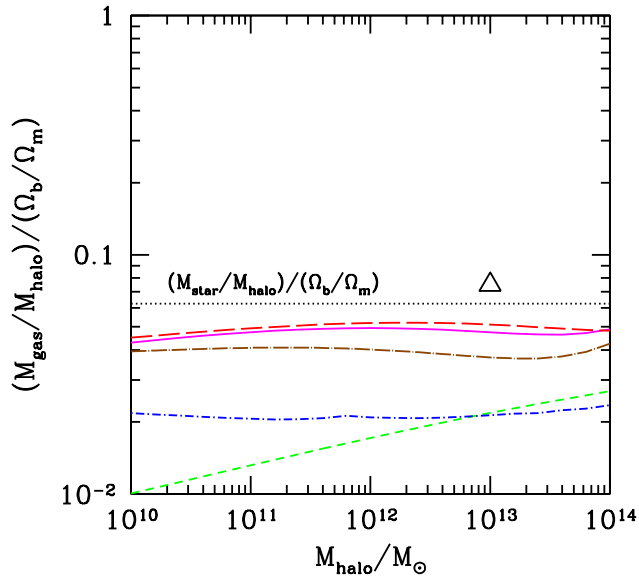}
\caption{\label{fig:mgas}
Fraction of halo baryons in the cold gas disk as a function of halo mass at $z=6$.  As in Fig. \ref{fig:rdist}, short-dashed (green), solid (magenta), and dot-short-dashed (blue) lines denote results for $\beta$-models with $\beta=1$, 0.1, and 0.01, while long-dashed (red) and dot-long-dashed (brown) curves indicate results for LSW models with $m=1$ and 0.02, respectively.  The gas mass inferred for HFLS3 by \citet{Riechers13} is noted by the triangle assuming a host halo mass of $10^{13}\,\msun$.  The horizontal, dotted line marks the typical fraction of halo baryons in stars at $z\sim6$ from \citet{Behroozi13}.
}
\end{center}
\end{figure}

The mass in cold gas in our disks is critical for the calculation of CO and [CII] lines since more gas will imply more emission.  We would naively expect the amount of cold gas to be approximately the same as the mass in stars, which in turn has been estimated via abundance matching to be roughly 1\% of the host halo mass \citep[e.g.,][]{Behroozi13}.  However, without making any additional assumptions, we can calculate the gas mass from our disk models by simply integrating the density required to ensure $Q=1$ at each radius.  Figure \ref{fig:mgas} shows the total gas masses at $z=6$ in our models for both LSW and $\beta$-model gas transport and different values of $m$ and $\beta$ as a function of halo mass.  The masses have been scaled by halo mass and the cosmic baryon fraction to give the fraction of halo baryons in the cold phase of the disk.

Our model results are not very sensitive to choice of model or parameter values---less than a factor of a few variation among all models considered---and are consistent with calculations of the mass of the cold neutral medium from the numerical simulations of \citet{Nagamine06b} and approximately equal to the stellar masses estimated by \citet{Behroozi13}.

The size, velocity, and star formation rate of HFLS3 are roughly consistent with its being hosted by a halo of mass just under $10^{13}\,\msun$.  \citet{Riechers13} find a total molecular mass of $(1\pm0.09)\times10^{11}\,\msun$ (assuming $\alpha_{\rm CO}=1$) and $2.0\times10^{10}\,\msun$ worth of atomic gas in the SMG.  Together these two components represent a total gas mass (denoted by the triangle in Fig. \ref{fig:mgas}) about 40\% larger than that in our $m=1$ LSW model for $10^{13}\,\msun$.  This is remarkably good agreement given the simplicity of our models and the uncertainties in $\alpha_{\rm CO}$.  Using the universal model for the CO to H$_2$ conversion factor in \citet{Narayanan12}, we estimate a value of $\alpha_{\rm CO}\approx0.84$ would be appropriate for HFLS3 given its CO luminosity and assuming solar metallicity gas.\footnote{We note, however, that \citet{Narayanan12} ignored the CMB as an observing background, which, while insignificant at $z=2$, is important at $z=6$ \citep{Obreschkow09b, MF13a, daCunha13b}.  As a result, the true value of $\alpha_{\rm CO}$ and the molecular mass associated with a given luminosity may be somewhat higher.}  Such a value brings the total inferred gas mass even closer to our estimate.  

\begin{figure*}
\begin{center}
\includegraphics[width=\textwidth,trim=0 260 0 0,clip]{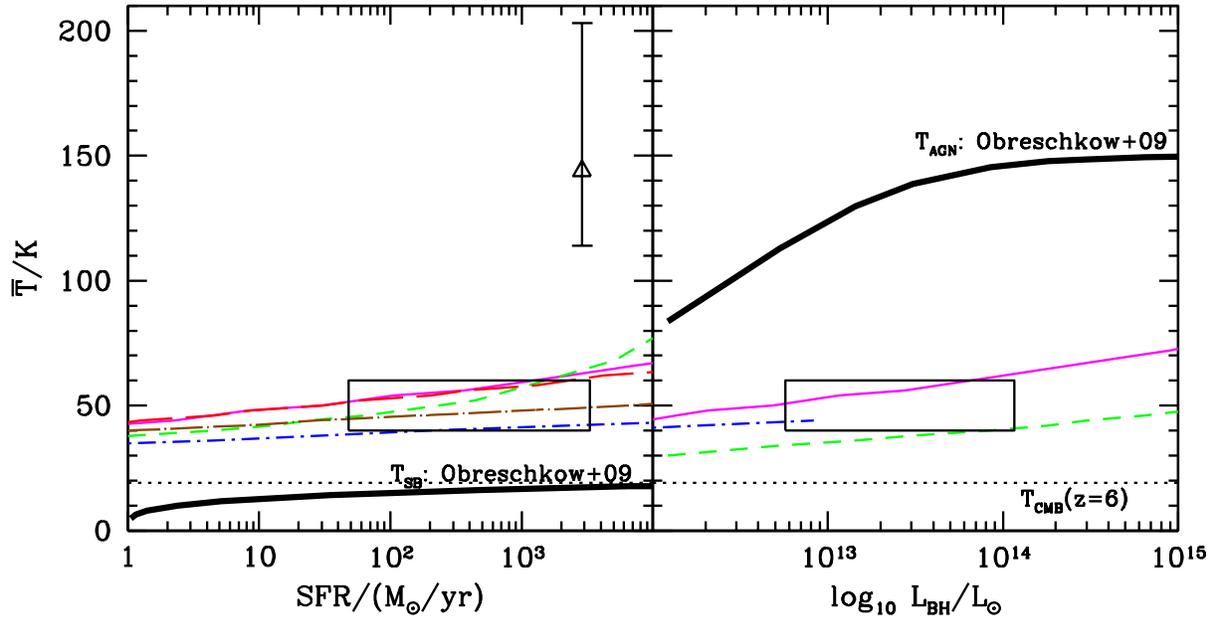}
\caption{\label{fig:Tave}
Average disk gas temperature as a function of star formation rate (left panel) and bolometric black hole luminosity (right panel) for our dusty disk models at $z=6$.  We define this average to be the temperature at which Eq. \ref{eq:Lcii} gives the same luminosity as our model if we assume thermal, optically thin gas.  As in Fig. \ref{fig:rdist}, short-dashed (green), solid (magenta), and dot-short-dashed (blue) lines denote results for $\beta$-models with $\beta=1$, 0.1, and 0.01, while long-dashed (red) and dot-long-dashed (brown) curves indicate results for LSW models with $m=1$ and 0.02, respectively.  Curves end at $10^{14}\,\msun$, the largest halo mass for which values were calculated.  Thick, black lines show the contributions from starbursts (left panel) and AGN (right panel) assumed in \citet{Obreschkow09b}.  The horizontal, dotted line indicates the CMB temperature of 19.11 K at $z=6$.  Boxes show the approximate range of MAMBO observations of quasar hosts, while the triangle (left panel) denotes the inferred kinetic temperature of HFLS3 from \citet{Riechers13}.
}
\end{center}
\end{figure*}

\subsection{Gas Temperatures}\label{sec:comp:temp}

Similar to the gas mass, the gas temperature is also an important ingredient in determining the luminosity of emission lines.  While the temperature varies as a function of radius throughout our model disks, a single value is often inferred observationally for the emitting gas in an entire galaxy.  Therefore, we calculate an equivalent temperature from our models for each halo mass and angular momentum transport model.  First, we calculate the [CII] luminosity assuming thermal, optically thin gas at the temperature prescribed by our disk model:
\begin{equation}\label{eq:Lcii_th}
\lcii=\int_0^{\rdisk}\!\! \frac{2\,{\rm \pi}\,r\,\left(1-\fco\right)\,{\rm C/H}}{m_{\rm p}}\,\frac{4\,{\rm e}^{-\hp\,\nu_{\rm [CII]}/k_{b}\,T}}{2+4\,{\rm e}^{-\hp\,\nu_{\rm [CII]}/k_{b}\,T}}\,A_{\rm [CII]}\,\hp\,\nu_{\rm [CII]}\,\Sgas\,dr,
\end{equation}
where $A_{\rm [CII]}=2.3\times10^{-6}\,{\rm s^{-1}}$, $\nu_{\rm [CII]}=1900.5\,{\rm GHz}$, the carbon abundance is ${\rm C/H}=1.5\times10^{-4}\,Z'$, $m_{\rm p}$ is the proton mass and where $\fco$, $T$, and $\Sgas$ are each functions of the galactocentric radius, $r$.  We then determine the temperature, $\bar{T}$, at which this result is reproduced by equation \ref{eq:Lcii}, taking the gas mass, $\mgas$, and the total fraction of gas containing CO, $\bar{f}_{\rm CO}$, from our disk model.  We plot this average temperature as a function of total star formation rate (left panel) and bolometric black hole luminosity (right panel) in Figure \ref{fig:Tave}.  

To compare $\bar{T}$ to observations, we note that the dust and gas are implicitly assumed to be strongly coupled at the same temperature in our model, and observationally, the dust temperature traces the gas temperature in most systems \citep[][but see discussion of HFLS3]{Malhotra01}.  Moreover, if ionized gas is a significant source of [CII] emission, then the observed dust temperature should include contain a contribution from these hotter pockets unless HII regions are completely devoid of dust.  The estimated temperature of the dust responsible for the strong 250 GHz continuum emission observed in roughly 30\% of $z=6$ SDSS quasars by the MAMBO survey \citep{Bertoldi03a, Petric03, Wang11b} is in the range of 40--60${\rm K}$ \citep[see also][]{Wang13}.  We fiducially set an approximate span of star formation rates for these systems from J0210$-$0456 and J1148$+$5251, the quasars with the lowest and highest measured values in our sample of quasars (though the undetermined values for J0129$-$0035 and J2329$-$0301 may turn out to be lower).  Similarly, the faintest and brightest bolometric black hole luminosities in our sample range from 0.53--11.6$\times 10^{13}\,{\rm \lsun}$ (from J0210$-$0456 and J1044$+$5251, respectively).  In Figure \ref{fig:Tave}, we compare this set of observed values to our model calculations of $\bar{T}$ and find good agreement with the quasar observations independent of the specifics of angular momentum transport.  Because our model does not allow the properties of the central emitting black hole to affect the gas in the outer parts of the galactic disk, this consistency implies that quasar heating of galactic gas---at least of the gas that dominates the [CII] emission---is minimal.  Note that in the right panel, we do not calculate $\bar{T}$ beyond $10^{14}\,{\rm \msun}$.

The agreement between our model and the observed quasars is in contrast to the discrepancy with regard to HFLS3 (triangle).  \citet{Riechers13} infer a much higher kinetic gas temperature---$144^{+59}_{-30}\,{\rm K}$---than predicted by any of our models.  However, the fact that the SED-derived dust temperature is so much lower than that of the gas---$56^{+9}_{-12}\,{\rm K}$---suggests that our coupled dust-gas model may be poor description of this system.

We further compare our results to the star burst and AGN contributions to the gas temperature assumed by \citet{Obreschkow09b}: $T_{\rm SB}=T^{\rm max}_{\rm SB}\,\left[\dot{\Sigma}_{\rm SF}/\left(\dot{\Sigma}_{\rm SF}+\dot{\Sigma}^{\rm c}_{\rm SF}\right)\right]^{1/4}$ and $T_{\rm AGN}=T^{\rm max}_{\rm AGN}\,\left[\dot{M}_{\rm BH}/\left(\dot{M}_{\rm BH}+\dot{M}^{\rm c}_{\rm BH}\right)\right]^{1/4}$, respectively, where $T^{\rm max}_{\rm SB}=60\,{\rm K}$, $\dot{\Sigma}^{\rm c}_{\rm SF}=500\,{\rm \msun/yr/kpc^2}$, $T^{\rm max}_{\rm SB}=150\,{\rm K}$, and $\dot{M}^{\rm c}_{\rm AGN}=10\,{\rm \msun/yr}$.  For this comparison, we calculate the average star formation rate surface density of a galaxy as $\dot{\Sigma}_{\rm SF}=\sfr/{\rm \pi}\,\rdisk^2$ and the black hole accretion rate as $\dot{M}_{\rm BH}=\lbh/(0.08\,c^2)$.\footnote{This is not quite equivalent to the procedure in \citet{Obreschkow09b}, where the radius used to compute $\dot{\Sigma}_{\rm SF}$ is the half-mass radius of the molecular gas rather than the disk radius.  However, the final temperature is not very sensitive to this difference.  The star formation rate surface density may rise higher than this average value in the inner regions of galaxy (both real and in our model), but this resolution is not achievable in the \citet{Obreschkow09b} semi-analytic prescription.}\!\! \footnote{A radiative efficiency of 8\% is appropriate for Schwarzchild black holes ignoring a general relativistic correction of order unity, but the efficiency may be as high as tens of percent for rotating black holes.  However, the final temperature is insensitive to these subtleties.}  $T_{\rm AGN}$ clearly overestimates the temperature of the gas and dust in these systems, while $T_{\rm SB}$ under-estimates it.  These deficiencies are likely symptomatic of over-extrapolation from fits at lower redshift.  However, we note that the kinetic temperature of HFLS3 may be reproduced by the \citet{Obreschkow09b} model if this system were, in fact, a quasar with $\lbh\gtrsim10^{13}\,\lsun$.

While our model considers in detail the emission from the PDRs of cold molecular clouds, recent works have attributed the [CII] emission at high-redshift to smoother, warmer gas components---$\gtrsim200\,{\rm K}$ \citep{Gong12, Vallini13}.  While these studies considered the emission from LBGs and LAEs rather than from quasars, their temperatures still exceed those found in the much larger systems.  This suggests that PDRs of molecular clouds are indeed the dominant source of [CII] emission at high redshift.

\section{Carbon Emission Lines}\label{sec:c}

In the previous section, we showed that our disk models provide reasonable descriptions for the observed systems considered in this work.  We are now in a position to calculate the amount of [CII] emission in our model galaxies.  In \S\ref{sec:c:cii}, we present results from the fiducial model described in \S\ref{sec:model} and compare them with observations.  Then, in \S\ref{sec:c:alt}, we consider changes to our base model suggested by the observations to improve agreement.  Finally, we will turn our attention the line luminosities of CO in\S\ref{sec:c:co}.

\subsection{The Fiducial [CII] Model}\label{sec:c:cii}
\begin{figure}
\begin{center}
\includegraphics[width=\columnwidth,trim=30 75 0 0,clip]{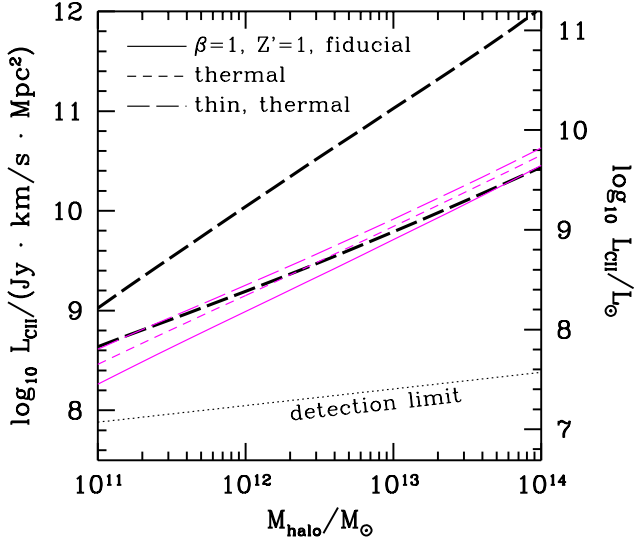}
\caption{\label{fig:cii-mhalo}
[CII] luminosity as a function of halo mass for our model quasar hosts.  The solid (magenta) curve shows results for a fiducial model of $\beta$-disks with $\beta=0.1$ and $Z'=1$, where the emission has been calculated using our modified version of the \citet{KT07} escape probability code.  The short-dashed and long-dashed curves explore the effects of cumulatively ({\emph{not}} alternative) assuming a thermal populated of states and optically thin gas.  Thick, black curves assume an optically thin, thermal population at a fixed temperature of 50 K and including either all disk gas (upper) or only gas containing CII (lower).  Finally, these models can be compared to the approximate 10 hour ALMA detection threshold for [CII] at $z=6$---extrapolated from the Cycle 0 configuration of \citet{Wang13}---denoted by the dotted curve.
}
\end{center}
\end{figure}

Combining the gas masses and temperatures described in \S\ref{sec:comp:mgas} and \S\ref{sec:comp:temp} with the physical properties of our molecular cloud PDRs, Figure \ref{fig:cii-mhalo} shows the resulting model [CII] luminosities for $z=6$ quasar hosts as a function of halo mass.  Here we assume $\beta=0.1$ and $Z'=1$ to demonstrate the effect of several different assumptions on the calculation.

The upper, thick, long-dashed line in the figure shows the simplest method for calculating the [CII] luminosity; the emission is derived from equation \ref{eq:Lcii} assuming thermal, optically thin gas with $\bar{f}_{\rm CO}=0$ and $\bar{T}=50\,{\rm K}$.  The choice of $\bar{f}_{\rm CO}=0$ implies that carbon everywhere in the galaxy is in the form of dissociated CII.  However, at the surface densities of halos larger than about $10^{11}\,\msun$, the gas is expected to be nearly all in molecular form \citep{Krumholz09b} and, at solar metallicity, nearly all of the carbon is in the form of CO \citep[][Eq. \ref{eq:fco}]{Wolfire10} with only a small fraction in CII.  Using $\bar{f}_{\rm CO}$---calculated properly according to our model via equations \ref{eq:fmol} and \ref{eq:fco}---results in the lower, thick, long-dashed curve in the figure and represents the largest effect toward accurately determining the amount of [CII] emission.  The effect is stronger at higher halo masses as the commensurately higher surface densities keep more of the carbon in the form of CO.  At low metallicities, there is much less dust in our model, which results in lower values of $\fco$, but the net [CII] emission ultimately decreases because, while $(1-\bar{f}_{\rm CO})$ asymptotically approaches unity, the total amount of carbon scales as $Z'$.  However, at high metallicities, the [CII] emission also decreases because $(1-\bar{f}_{\rm CO})$ decreases faster than the carbon abundance increases.

Continuing to build up the complexity of the calculation in Figure \ref{fig:cii-mhalo}, the thin, long-dashed (magenta) curve shows the emission determined using equation \ref{eq:Lcii_th}; the gas is still assumed to be thermal and optically thin, but now the gas temperature is allowed to varying according to our disk model.  Note that the approximate agreement between equations \ref{eq:Lcii_th} and \ref{eq:Lcii} for $\bar{T}=50\,{\rm K}$ was previously indicated by the flatness of the curves in Figure \ref{fig:Tave}.  The thin, short-dashed (magenta) curve in Figure \ref{fig:cii-mhalo} further improves the calculation of the [CII] emission by relaxing the assumption of optically thin gas.  This additional complication does not significantly decrease the predicted emission.  Finally, the full calculation is denoted by the thin, solid (magenta) line, which shows a very similar prediction when the assumption of thermal emission is removed---less than a factor of two decrease in $10^{11}\,\msun$ halos.  Thus, Figure \ref{fig:cii-mhalo} demonstrates that our model luminosity can be approximated---at the large halo masses considered here and for a $\beta=0.1$ model---by assuming thermal, optically thin gas at a fixed temperature of $\sim 50\,{\rm K}$.  These assumptions get worse as halo mass decreases, and care should be taken when considering the high-redshift LBGs hosted by smaller systems in future work.

\begin{figure*}
\begin{center}
\includegraphics[width=\textwidth,trim=0 40 0 60,clip]{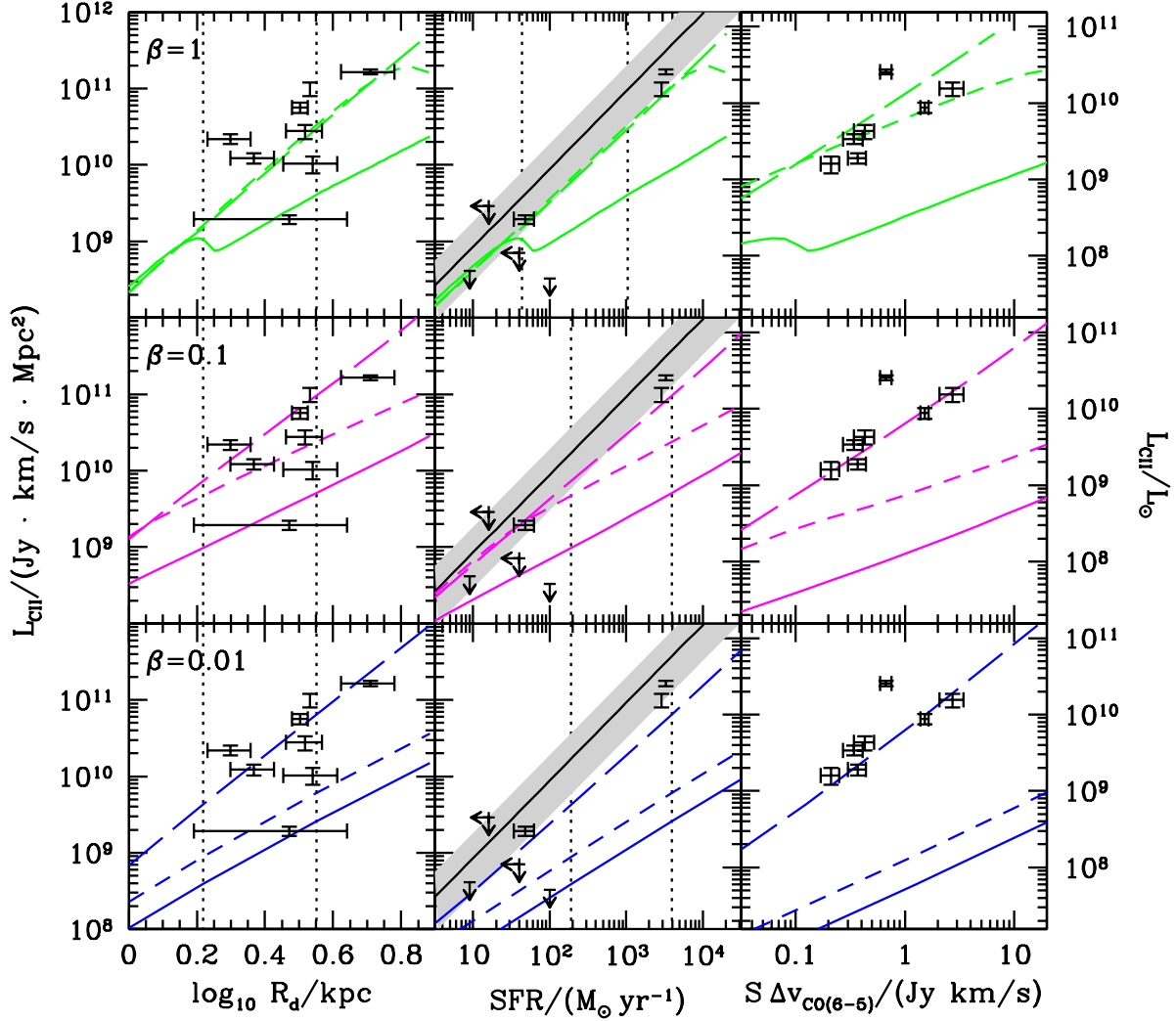}
\caption{\label{fig:cii1}
[CII] luminosity as a function of $\rdisk$ (left column), $\vcirc$ (middle column), and star formation rate (right column) assuming solar metallicity ($Z'=1$).  Top, middle, and bottom rows show our model results for dusty, $\beta$-disks with $\beta=1$, 0.1, 0.01, respectively.  In each panel, we compare results for Models 1 (solid curves), 2 (short-dashed), and 3 (long-dashed) with the points representing the data listed in Tables \ref{tab:disk} and \ref{tab:lum} (we have removed the different point types for clarity).  Vertical dotted lines indicate disk radii and star formation rates corresponding to $10^{12}$ and $10^{13}\,\msun$; since the relationship between CO(6--5) flux and halo mass changes for different Models, these lines do not appear in the left-hand panels.  In the central column, the solid, black line and shaded region denote the observed mean and scatter of the local relation initially compiled by \citet{deLooze11} and recalculated by \citet{Ouchi13}.
}
\end{center}
\end{figure*}

\begin{figure*}
\begin{center}
\includegraphics[width=\textwidth,trim=0 180 0 60,clip]{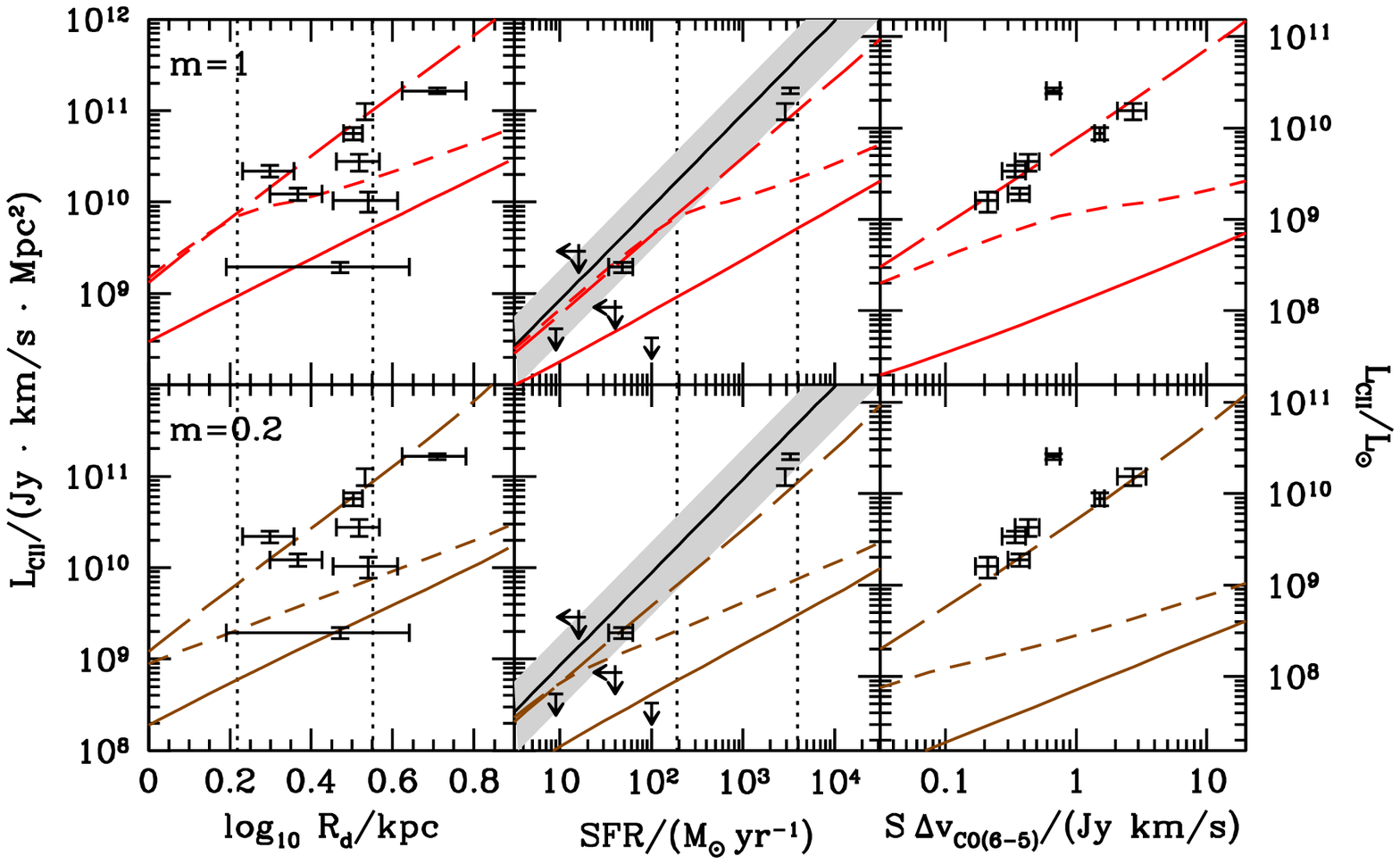}
\caption{\label{fig:cii2}
Same as Fig. \ref{fig:cii1} except for LSW models with $m=1$ and 0.2 in the top and bottom rows, respectively.
}
\end{center}
\end{figure*}

In Figures \ref{fig:cii1} and \ref{fig:cii2}, we show the [CII] luminosity assuming solar metallicity ($Z'=1$) as a function of disk radius, total star formation rate, and flux in the CO(6--5) line for a range of angular momentum transport parameter-space: $\beta$-models with $\beta=0.01$, 0.1, and 1 (Fig. \ref{fig:cii1}) and LSW models with $m=0.2$ and 1 (Fig. \ref{fig:cii2}).  As discussed in \S\ref{sec:comp:lbh}, we expect the $\beta$-models of Figure \ref{fig:cii1} to be more descriptive of quasar hosts with $\beta$ close to 0.1 (middle row) if we neglect quasar outflows and 1 (top row) if outflows are significant.  On the other hand, the LSW models of Figure \ref{fig:cii2} are likely more representative of typical galaxies such as LBGs and LAEs, though perhaps at lower metallicities.

In both figures, solid curves show our fiducial calculations as described in \S\ref{sec:model}, which we denote ``Model 1."  We find that Model 1 under-predicts the observed [CII] luminosities in our sample by approximately an order-of-magnitude on average at fixed $\rdisk$ or star formation rate.  At fixed $\Sdvco$, the agreement is even worse, but we expect that this is due to a corresponding over-production of CO(6--5) emission (see \S\ref{sec:c:co}).  We anticipated this degree of discrepancy between our models and the observations in \S\ref{sec:intro:cii}.  Simply increasing the metallicity beyond $Z_{\odot}$ will produce more total carbon, but the implied increase in dust content will put a higher fraction of the carbon in the form of CO.  In \S\ref{sec:c:alt}, we will consider two interesting departures from our fiducial model suggested by the observations to improve agreement; these alternatives are indicated by short- and long-dashed lines in the figures.

\begin{table}
\begin{center}
\begin{tabular}{lcc}
\hline
\multicolumn{3}{c}{Molecular Gas Models} \\ 
\hline
model name & implementation & possible physical cause \\
\hline
Model 1 & described in \S\ref{sec:model} & fiducial case \\
\hline
Model 2 & multiply $\sfrff/\tff$ & higher star \\
& by factor of 10 &  formation efficiency \\
\hline
Model 3 & divide $\fco/\fmol$ & lower dust-to- \\
& by factor of 10 & metals ratio \\
\hline
\end{tabular}
\caption{Fiducial and alternative molecular gas models considered in \S\ref{sec:c}.  Solid, short-dashed, and long-dashed curves in Figures \ref{fig:cii1}, \ref{fig:cii2}, and \ref{fig:lbh-cii} denote results from Models 1, 2, and 3, respectively.
}
\label{tab:mod}
\end{center}
\end{table}

In the central columns of Figures \ref{fig:cii1} and \ref{fig:cii2}, we compare our sample and Model 1 results to the local scaling relation of [CII] luminosity with star formation rate \citep{deLooze11, Ouchi13}.  The quasar measurements and that of HFLS3 are consistently below the average local relation but still within the observed scatter.  However, \citet{Ouchi13} recently placed a 3$\sigma$ upper limit on the [CII] luminosity of {\it{Himiko}}---$\lcii<3\times10^{8}\,{\rm Jy km/s Mpc^2}$---and determined an estimated star formation rate of $100\,{\rm \msun/yr}$, which place the LAE significantly below the local relation.  These authors suggested that such an offset implies a metallicity for {\it{Himiko}} below 10\% of solar.  If Model 1 is an accurate description of LAEs---despite its deficiencies with respect to quasars---then the metallicity of {\it{Himiko}} could be as high as solar if the inflow rate of gas is quite low $\beta\lesssim0.01$ (consider the bottom, central panel of Fig. \ref{fig:cii1})  or $m\lesssim0.2$ (bottom, central panel of Fig. \ref{fig:cii2}).  However, we showed in \citet{MF13a} that a consistent picture of the molecular fraction in chemical equilibrium in typical $z\sim6$ galaxies implies gas-phase metallicities of only a few percent of solar.  Thus, the likeliest scenario is that the metallicity of {\it{Himiko}} is indeed very low and that Model 1 does a poor job of representing the [CII] emission across our whole sample.

\subsection{Alternative Molecular Gas Models}\label{sec:c:alt}

Given that we roughly predict the correct gas masses, star formation rates, and temperatures of these systems, the disagreement between observations and Model 1 must lie in the ratio of [CII]-traced to CO-traced gas.  Figure \ref{fig:cii-mhalo} indicated that we can increase the amount of [CII] emission by shifting more of the carbon from CO into dissociated CII by decreasing either $\fmol$ or $\fco/\fmol$ or both.  A decrease in $\fmol$ is represented in our model by an increase in the efficiency of star formation from molecular gas---thus maintaining the $\dSstar$ required at each radius to ensure $Q=1$---but may physically be a result of the destruction of molecular clouds before the equilibrium value of $\fmol$ is reached \citep{MLG12} and star formation occurring in atomic gas \citep{Krumholz12}.  We, thus, define ``Model 2" as one in which $\sfrff/\tff$ from equation \ref{eq:fmol} increased by a factor of 10 (short-dashed curves in Figs. \ref{fig:cii1} and \ref{fig:cii1}).  We further consider a model in which we instead manipulate the fraction of gas containing CO directly, dividing $\fco/\fmol$ given in equation \ref{eq:fco} by a factor of 10, and denote this variation as ``Model 3" (long-dashed curves).  However, it is worth keeping in mind that both effects could be happening simultaneously to some degree.  Table \ref{tab:mod} summarizes the three different molecular gas models we consider in this work.

As shown in Figures \ref{fig:cii1} and \ref{fig:cii2}, the predicted [CII] emission increases substantially in Models 2 and 3.  For Model 2, the increase in the [CII] luminosity is sufficient to explain the observations only if gas is transported very quickly through the disk with $\beta=1$.  On the other hand, Model 3 reproduces the observations more generically; the relationships between [CII] emission and either star formation rate or $\Sdvco$ are more robust to changes in the angular momentum transport mechanism or parameter values.  In this case, with respect to star formation rate, we find, 
\begin{equation}\label{eq:lcii-sfr}
\lcii=5\,\times10^{8}\,\lsun\,\left(\frac{{\rm SFR}}{100\,{\rm \msun/yr}}\right)^{0.9}.
\end{equation}
As a function of CO flux, all but one of the models give results well-presented by
\begin{equation}\label{eq:lcii-lco6}
\lcii=6\,\times10^{9}\,\lsun\,\left(\frac{\Sdvco}{{\rm Jy\,km/s}}\right)^{0.9}
\end{equation}
with the values for $\beta=1$ being about 0.3 dex brighter.  

As discussed in \S\ref{sec:c:cii}, the upper limits on the LAE {\it Himiko} are inconsistent with Models 2 and 3 for $Z'=1$.  Similarly, HCM 6A (central panels; SFR$=9\,{\rm \msun/yr}$; $\lcii<4.1\times10^{8}\,{\rm Jy km/s Mpc^2}$) may only be in marginal agreement with Models 2 and 3 at solar metallicity (consider the bottom, central panel of Fig. \ref{fig:cii2}).  A stronger disagreement would result if the star formation rate for this LAE is significantly higher than that estimated from its UV continuum.  However, this tension may simply indicate a metallicity for HCM 6A somewhat less than solar.

\begin{figure}
\begin{center}
\includegraphics[width=\columnwidth]{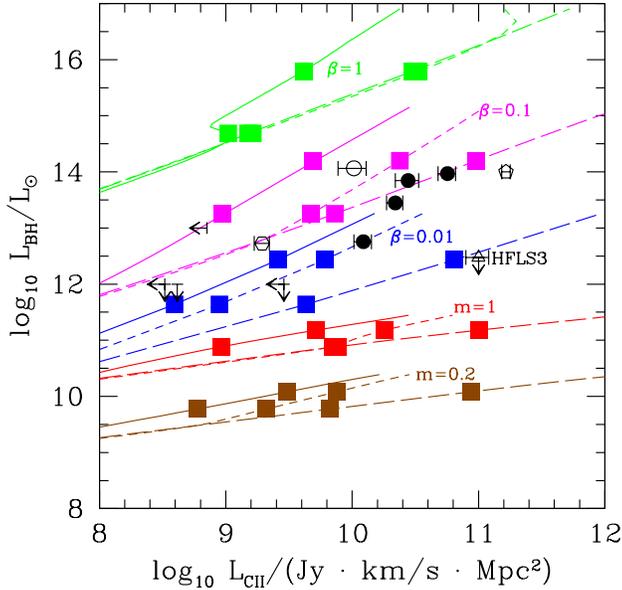}
\caption{\label{fig:lbh-cii}
Bolometric black hole luminosity as a function of [CII] luminosity.  This figure combines results from Fig. \ref{fig:lbh}, \ref{fig:cii1}, and \ref{fig:cii2}.  As in Fig. \ref{fig:lbh}, green, magenta, and blue lines correspond to $\beta$-models with $\beta=1$, 0.1, and 0.01, while red and brown curves denote LSW models with $m=1$ and 0.2, respectively.  Solid, short-dashed, and long-dashed curves indicate results for Models 1, 2, and 3, respectively, and filled squares denote $10^{12}$ and $10^{13}\,\msun$ for each model.  Lines end at $10^{14}\,\msun$, the largest halo mass for which values were calculated.  As in Fig. \ref{fig:disk}, circles represent the observed quasar hosts of \citet{Wang13}, the pentagon is J1148$+$5251, the hexagon is J0210$-$0456, the triangle is the SMG HFLS3, and the limits correspond to J2329$-$0301 and the LAEs.  Also as before, filled points indicate observations for which ${\rm sin} i$ is available, though no inclination correction is relevant for this figure.
}
\end{center}
\end{figure}

In Figure \ref{fig:lbh-cii}, we combine our [CII] calculations from Figures \ref{fig:cii1} and \ref{fig:cii2} with our black hole accretion rate results from Figure \ref{fig:lbh}.  We plot [CII] luminosity versus bolometric black hole luminosity for Models 1, 2, and 3.  Model 3 with $\beta \approx 0.1$ clearly does the best job of describing the observed quasars for reasonable halo masses ($10^{12}$--$10^{13}\,\msun$).  Here, HFLS3 seems more in tension with $\beta$ even as low as 0.01.  If SMGs in halos comparable to those of quasars are actually better-described by a different mode of angular momentum transport (in this case, by LSW disks), it may have implications for the quasar duty cycle and the ability of systems to form supermassive black holes by $z\sim7$ \citep{Mortlock11}.  

As in \S\ref{sec:comp:lbh}, we again find that the quasar J2329$-$0301 (SFR$<40\,{\rm \msun/yr}$; $\lbh=10^{13}\,\lsun$; $\lcii<7.9\times10^{8}\,{\rm Jy km/s Mpc^2}$) is consistent with our models with no hint of the uniqueness ascribed to it by \citet{Willott13}, though Figure \ref{fig:lbh-cii} suggests it may have a value of $\beta$ slightly above 0.1 or non-negligible mass drain from either a quasar wind or a growing bulge.  While this consistency may hardly be surprising given that only upper limits exist on both the star formation rate and [CII] luminosity for this object, it is worth noting that a star formation rate above the $40\,{\rm \msun/yr}$ upper limit would be inconsistent with the upper limit on its [CII] flux in Models 2 and 3 (central panels in Fig. \ref{fig:cii1}), particularly for $\beta>0.1$.

\subsection{CO Line Ratios}\label{sec:c:co}

\begin{figure}
\begin{center}
\includegraphics[width=\columnwidth]{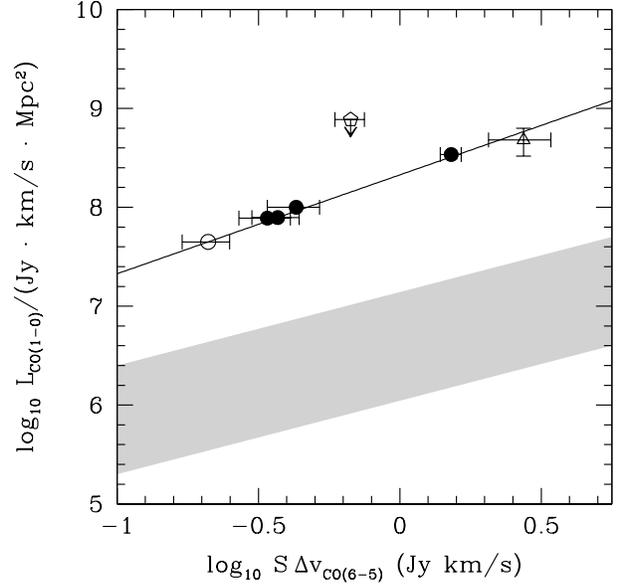}
\caption{\label{fig:co}
The CO(6--5) velocity-integrated flux versus CO(1--0) luminosity.  The gray, shaded region shows the range predicted by our model when both the angular momentum transport mechanism and the choice of Models 1, 2, or 3 is varied.  As in Fig. \ref{fig:disk}, circles represent the observed quasar hosts of \citet{Wang13} with CO(1--0) values estimated from the J1148$+$5251 excitation ladder of \citet{Riechers09}---the solid, black line indicates the enforced proportional relationship between the emission from the two emission lines.  The pentagon is J1148$+$5251 itself and the triangle is the SMG HFLS3.  Also as before, filled points indicate observations for which ${\rm sin}\,i$ is available, though no inclination correction is relevant for this figure.
}
\end{center}
\end{figure}

In this section, we turn our attention from the [CII] line to those of CO.  While we considered the line luminosities of CO in typical LBGs during reionization in previous work \citep{MF13a}, we revisit it here for more extreme systems and focus particularly on the ratio of the $J=6\rightarrow5$ and $J=1\rightarrow0$ lines.  Figure \ref{fig:co} shows the CO(1--0) luminosity as a function of CO(6--5) velocity-integrated flux for our models compared to observations.  Circles mark the \citet{Wang13} sample of $z\sim6$ quasars.  For these objects, CO(1--0) luminosity values are not directly observed but, rather, extrapolated from the excitation ladder of J1148$+$5251 \citep[][pentagon]{Riechers09} by assuming a proportional relationship with CO(6--5) as indicated by the solid line.  The detected CO lines from HFLS3 (triangle) are consistent with this extrapolation.  For convenience, we denote our models by a shaded region that encompasses the range for a variety of angular momentum transport parameters and for Models 1, 2, and 3.  All of our models seem roughly to agree with the expected proportional relationship between emission from the two CO lines, though the shaded region smears out some scatter in the slope of our model relation among different parameter sets.  This suggests that the excitation states of the lines do not vary strongly with halo mass but, rather, that larger halo masses produce brighter lines simply because their hosted galaxies contain more gas.  

However, we fail to reproduce the ratio of these two lines, predicting much more CO(6--5) flux for a given CO(1--0) luminosity.  If we assume thermalized and optically thin lines at an excitation temperature of, $T$, the ratio between the luminosities of any two CO lines, represented by upper-state rotational quantum numbers $J$ and $K$, is
\begin{eqnarray}\label{eq:Lrat}
\frac{L_{J \rightarrow J-1}}{L_{K \rightarrow K-1}}&=&\frac{J}{K}\,\frac{A_{J}}{A_{K}}\,\frac{n_{J}}{n_{K}}\nonumber\\ 
&=&\frac{J\,(2\,J+1)}{K\,(2\,K+1)}\,\frac{A_{J}}{A_{K}}\,{\rm e}^{-\frac{(J-K)\,(J+K+1)\,\hp\,\nu_{\rm CO}}{2\,k_{b}\,T}}.
\end{eqnarray}
For CO, $A_{J=1}=7.203\times10^{-8}\,{\rm s^{-1}}$ and $A_{J=6}=2.137\times10^{-5}\,{\rm s^{-1}}$, while $\nu_{\rm CO}=115.3\,{\rm GHz}$.  For $J=6$, $K=1$, and $T=50\,{\rm K}$, equation \ref{eq:Lrat} gives about 800, roughly a factor of four higher than the \citet{Wang13} value of $\lcoo/\lco \approx 170$.  Assuming a temperature of $40\,{\rm K}$ results in a ratio of roughly 500.  Of course, the lines are likely not both thermally populated and perhaps not optically thin.  Of the two, the $J=6$ state is likely to be the most subthermal since its critical density is on the order of $10^5\,{\rm cm^{-3}}$.  Subthermal suppression should be a bigger effect than any optical thickness of the CO(1--0) line and will likely decrease the expected $\lcoo/\lco$ from the value approximated in equation \ref{eq:Lrat}.  Moreover, subtraction of the CMB background will not affect both lines equally; the CMB is brighter in the CO(6--5) line than in the CO(1--0) line, yet subtraction of the same background will affect optically thick lines more than optically thin ones.  Despite achieving roughly correct gas temperatures, our model over-produces the CO(6--5) flux for a given CO(1--0) luminosity, predicting $\lcoo/\lco \sim 10^3$, close to the thermal, optically thin value of equation \ref{eq:Lrat}.  Additionally, our models produce about the same ratio independent of angular momentum transport mechanism or Model (i.e., any model variation summarized in Tables \ref{tab:amt} or \ref{tab:mod}).\footnote{Large deviations among model results occur only where our CMB subtraction implies that a line is seen in absorption rather than emission.  However, as in \citet{MF13a}, we caution against taking these values too seriously since our model more accurately predicts the intrinsic emission and the amount of absorption separately than it does the difference between the two when similarly valued.}  
The discrepancy implies that, inside real galaxies, the CO(6--5) line is more sub-thermally populated than predicted by our models.  The likely culprit is the high level of turbulent support (and associated density fluctuations) within our model's molecular clouds.

To see how turbulently-generated density fluctuations affect the thermalization of the CO(6--5) line in our model, consider the simplified treatment of a $10^{13}\,\msun$ halo from \S\ref{sec:intro:cii} hosting a galaxy with a surface gas density of $4200\,{\rm \msun/pc^2}$.  The Jeans mass at the disk radius for this surface density is $3\times10^9\,\msun$.  Clouds of this mass require a turbulent velocity dispersion of ~100 km/s to achieve a virial ratio of $\alpha=2$, and the resulting thermal sound speed at 50 K is ~0.05 km/s.  Thus, the 1D thermal Mach number is ~2000, about a factor of 20 larger than the Mach numbers in strong, local starbursts.  Since these objects have the same temperatures, the high Mach number results directly from the large cloud masses and high surface densities.  As a result, the mass-weighted median density is a factor of 1700 larger than the mean density.  Therefore, even though the critical density for thermalization of the CO(6--5) line is nearly a factor of 100 larger than that of CO(1--0), the high median density of the gas ensures that both lines are roughly thermal throughout our model disks.  In \citet{MF13a}, we showed how including the effect of such turbulent clumps may have important effects for the observability of high excitation CO lines in more typical galaxies at high redshift.

In contrast to these theoretical expectations, observers have found subthermally excited CO(6--5) lines in two systems, as indicated by the mean molecular densities found for J1148$+$5251 \citep{Riechers09} and HFLS3 \citep{Riechers13}: $n_{\rm H_2}=10^{4.2}\,{\rm cm^{-3}}$ and $n_{\rm H_2}=10^{3.8^{+0.28}_{-0.17}}\,{\rm cm^{-3}}$, respectively.  These authors fit the CO excitation using temperature and a single ``effective" density as free parameters.  However, following the elementary assumptions of our model, the median density of our clouds is much higher than these inferred values, owing to the aforementioned turbulent fragmentation in the disk.  Resolving the disagreement with observations will require changing the model of ISM physics to produce lower turbulent Mach numbers and hence lower median densities.

\section{Discussion \& Conclusions}\label{sec:discussion}

One of the main goals of this work is to understand whether our galaxy formation framework is an adequate description of observed LAEs, SMGs, and quasar hosts during the epoch of reionization.  In \S\ref{sec:comp:struc}, \S\ref{sec:comp:lbh}, \S\ref{sec:comp:mgas}, and \S\ref{sec:comp:temp}, we showed that the relationships among radii, circular velocities, star formation rates, black hole luminosities, gas masses, and temperatures are all reasonably well-captured by our model.  That we can come anywhere close to reproducing these properties with our idealized disk models is impressive, particularly with regard to quasar hosts, which are often thought to be built up in a succession of very rapid mergers \citep[e.g.,][]{Li07} since such mergers may provide the gravitational torques necessary to channel gas into the center of the disks.  We find that most of the quasar hosts at $z=6$ are well-described by $\beta$-disks with $\beta=0.1$ living inside $\sim10^{12}$--$10^{13}\,\msun$ halos with no direct feedback from the quasar onto star formation in the galactic disk.  We point out that the relatively low star formation rates of the CFHQS quasars J0210$-$0456 and J2329$-$0301 are completely consistent with this lack of feedback.  Our analysis ignores the effect of quasar outflows, which have been observed in some of these systems \citep{Maiolino12, Valiante12}.  Such outflows may imply that a higher value of $\beta$ is appropriate for the gas flows in these objects.  In this case, the extra gas that would have accreted onto the central black hole is instead expelled.  However, since our model reproduces the dependencies of $\lbh$ on $\rdisk$, $\vcirc$, and star formation rate while ignoring quasar winds, the inclusion of such outflows and the dependence of the outflow rate on quasar properties is necessarily highly constrained.  These constraints may prove an interesting avenue toward theoretically modeling quasar outflows in future work.  Further, we use fits to our model results (Eq. \ref{eq:fits}) to make predictions for the star formation rates of the \citet{Wang13} quasar hosts in advance of observational determinations from rest-frame UV (see Table \ref{tab:disk}).

The SMG HFLS3 is an interesting case for our models since it has a large radius, line width, and star formation rate but no apparent AGN activity.  If the data constrain the central bolometric black hole luminosity in HFLS3 to be less than $3\times10^{12}\,\lsun$---somewhat lower than could be observed in the deepest SDSS stripe based on its UV continuum component---the object would be only marginally consistent with a $\beta$-disk for $\beta=0.01$.  However, within the framework of our model, the large radius, broad line width, high star formation rate, and low AGN activity of HFLS3 could easily be explained by any of our LSW disks inside a halo just under $10^{13}\,\msun$.  

However, our fiducial treatment of molecular gas is not completely consistent with observations.  For example, we have difficulty reproducing the molecular gas fraction of HFLS3.  Our method of calculating $\fmol$ from equation \ref{eq:fmol} in \citet{MF13a} predicts an average molecular fraction of more than 99\% for this object using parameters consistent with its other observables.  Yet, given its inferred atomic gas content, this system has an average molecular fraction of only about 80\%.  The measurement of atomic gas via the observed [CII] emission could be an over-estimate if the signal were dominated by dissociated carbon in otherwise molecular gas, but this creates a new problem since our model also predicts $\fco/\fmol>0.97$, implying that nearly all of the molecular gas contains carbon in the form of CO.  We seem to under-estimate the amount of CII in this system, but whether this is because we over-estimate the molecular fraction, $\fmol$, or the fraction of un-dissociated molecular gas, $\fco/\fmol$, is unclear.  Regardless of the reason, this deficiency in our base model (Model 1) is related to our under-prediction of the [CII] luminosity compared to all of the detections we considered.  While our model does ignore the contribution from ionized gas, we argued that PDRs of molecular clouds are the dominant source of [CII] emission at these high redshifts.

Figures \ref{fig:cii1} and \ref{fig:cii2} also showed that our base model over-predicts the CO(6--5) fluxes of these systems if they are indeed associated with halos in the $10^{12}$--$10^{13}\,\msun$ range.  The combination of under-predicted [CII] and over-predicted CO again suggests that our models do not properly calculate $\fco$ in these systems.  Models 2 and 3 are two {\it{ad hoc}} variations to our base model (Model 1) that try to improve this deficiency.  While there are no [CII] or CO detections of $z\sim6$ LBGs, if our model similarly over-predicts $\fco$ in these systems, the prospects for observing them in CO would be even worse than suggested by \citet{MF13a}.

In Model 2, we increase the star formation rate per dynamical time, $\sfrff$, by a factor of ten (see summary in Table \ref{tab:mod}).  This efficiency is used to calculate the molecular fraction via equation \ref{eq:fmol}.  As discussed in \citet{MF13a}, we also set $\fmol=1$ when equation \ref{eq:fmol} would otherwise predict a value greater than unity.  This threshold, therefore, already effectively represents an increase in $\sfrff$ above the value predicted by \citet{Krumholz09b}.  The further increase to $\sfrff$ in Model 2 reduces the molecular fraction required to produce a given star formation rate and reduces the frequency with which the $\fmol=1$ cutoff comes into effect.  Physically, this change in the star formation efficiency may arise from the somewhat different physical properties expected in high-redshift molecular clouds, such as their very high surface densities or short free-fall times.  Alternatively, star formation in atomic gas \citep{Krumholz12} would provide the same ultimate result.  A contribution from HII regions may also work in a similar way by reducing the amount of molecular gas, albeit at higher temperatures.  Regardless of the physical mechanism, we assume that changing $\sfrff$ in this alternate model does not affect $\fco/\fmol$---calculated via equation \ref{eq:fco}---but ultimately affects $\fco$.  The magnitude of the increase in $\sfrff$ that we selected for Model 2 is somewhat arbitrary and was not chosen to produce agreement with the observations but rather to demonstrate the effect of the modification.  Nevertheless, we can ask whether this amount of increase in the star formation efficiency is sufficient to reproduce the observations.  We find that Model 2 changes the results in low-mass halos more than in high-mass ones.  This is expected since high-mass halos are more effected by the $\fmol=1$ cutoff which essentially nullifies the effect of increasing $\sfrff$ on $\fmol$.  The specific increase in Model 2 is only sufficient to explain the [CII] observations if $\beta=1$.  To further match the relationships among $\lbh$, $\rdisk$, $\vcirc$, and star formation rate in the quasar hosts, one must additionally invoke quasar outflows to reduce the accretion rate of the central black hole and the resulting luminosity.  For HFLS3, however, this model appears to fail---despite predicting a total molecular fraction of 80\%---unless $\sfrff$ is still higher than we set here, since the disk parameters that produce the correct [CII] luminosity over-predict that of the central black hole even with winds.

In Model 3, on the other hand, we create more CII by dissociating more of the carbon residing in otherwise molecular gas.  Here, we decrease $\fco/\fmol$ by a factor of ten while leaving the molecular fraction, $\fmol$, unchanged.  Physically, this scenario may result from a lower dust-to-metals ratio at high redshift so that the same metallicity produces less dust-shielding against CO dissociation.  Again, the magnitude of this change is somewhat arbitrary and set primarily for demonstration purposes.  Nevertheless, we find that Model 3 produces [CII] luminosity as a function of disk radius, star formation rate, or CO(6--5) flux consistent with observations independent of angular momentum transport mechanism.  Thus, unlike in Model 2, we can simultaneously reproduce both the [CII] luminosity and low central black hole accretion rate of HFLS3.

A second major issue revealed by our work is that, while Models 2 and 3 seem to do a better job of producing the observed [CII] emission than does Model 1, none of these scenarios adequately describes the CO excitation ladder of $z\sim6$ systems.  Our model does yield a roughly constant value of $\lcoo/\lco$ with luminosity suggesting that the fluxes scale with gas mass, but our low value of the ratio compared to observations likely results from an over-estimate of the Mach number in our molecular clouds and a correspondingly more thermalized CO(6--5) population than observed.  We can reproduce the correct ratio of CO(6--5)/CO(1--0) in our models if the Mach numbers are about a factor of 12 lower than we estimate.  Since our temperatures are consistent with observations, lower Mach numbers require lower turbulent velocity dispersions.  While this could be effected by a lower virial ratio, it is unclear how to physically interpret the necessary value of $\alpha\sim0.01\ll1$.  Further, this scenario conflicts with the results of \citet{Mashian13}.  These authors fit the velocity gradient and gas density for the $z\sim5$ SMG HDF 850.1 under different virialization assumptions and concluded that high-redshift molecular clouds are likely un-virialized, i.e., $\alpha>1$, though they did not take into account the effect of the velocity gradient on the clumpiness of the gas in the case that the support is provided by turbulence.  Alternatively, a lower velocity dispersion could result from smaller clouds.  While we assume $M_{\rm cl}=M_{\rm Jeans}$ as a function of radius, where our estimate of the Jeans mass already takes into account the self-gravity of the gas disk, a large contribution to the self-gravity from stars could lower the Jeans mass further.  Clouds might also have smaller masses if they form as a result of turbulent rather than gravitational fragmentation and can achieve a dissociative equilibrium before they can be smoothed out by the Jeans stability.  Finally, we note that, while the magnitude of the effect is unclear, fewer turbulent clumps could also result in more dissociation of CO into CII and boost the [CII] luminosity.

This study reveals two puzzles in relating recent observations to existing ISM theory---namely, reproducing the [CII] emission and the ratio CO(6--5)/CO(1--0)---and suggests some schematic ways to improve agreement.  Clearly, as samples of $z\gtrsim6$ systems with well-determined atomic and molecular emission continue grow, more theoretical work will be needed to understand the small scale physics of high-redshift molecular clouds and appreciate the differences between those at low-redshift.

\section{Acknowledgements}

We thank Jean Turner, Mark Krumholz, Chris Carilli, Dominik Riechers, Matt Malkan, and Desika Narayanan for helpful discussions and suggestions.  This research was partially supported by the David and Lucile Packard Foundation.

\bibliography{ms.bib}

\end{document}